\documentclass[aps,prl,reprint,superscriptaddress,longbibliography]{revtex4-2}

\usepackage[T1]{fontenc}
\usepackage[utf8]{inputenc}
\usepackage{amsmath,amssymb,amsfonts,mathtools,bm}
\usepackage{braket}
\usepackage{graphicx}
\usepackage[table]{xcolor}
\usepackage{booktabs,array}
\usepackage{microtype}
\usepackage{hyperref}
\usepackage[ruled,vlined,linesnumbered]{algorithm2e}

\definecolor{PDAblue}{HTML}{173F5F}
\definecolor{PDAaccent}{HTML}{2A6F97}
\definecolor{PDAlight}{HTML}{EAF2F8}
\definecolor{PDAstripe}{HTML}{F7FAFC}
\definecolor{PDArule}{HTML}{9FB7C8}
\definecolor{AlgBlue}{HTML}{245A8D}
\definecolor{AlgTeal}{HTML}{168C8C}
\definecolor{AlgGray}{HTML}{5F6B73}
\arrayrulecolor{PDArule}

\newcommand{\ii}{\mathrm{i}}
\newcommand{\ee}{\mathrm{e}}
\newcommand{\Id}{\mathbf I}
\newcommand{\RR}{\mathbf R}

\newcommand{\safeincludegraphics}[2][]{%
  \IfFileExists{#2}{\includegraphics[#1]{#2}}{%
  \fbox{\parbox[c][3.0cm][c]{0.92\linewidth}{\centering
  Figure placeholder:\\[1mm]\texttt{\detokenize{#2}}}}}%
}

\hypersetup{
  colorlinks=true,
  linkcolor=blue!55!black,
  citecolor=blue!55!black,
  urlcolor=blue!55!black
}

\begin{document}

\title{Universal Quantum Criticality from a Gaussian Variational Manifold}

\author{Reyhaneh Khasseh}
\thanks{These authors contributed equally to this work.}
\affiliation{Theoretical Physics III, Center for Electronic Correlations and Magnetism, Institute of Physics, University of Augsburg, D-86135 Augsburg, Germany}

\author{Fabian Ballar Trigueros}
\thanks{These authors contributed equally to this work.}
\affiliation{Theoretical Physics III, Center for Electronic Correlations and Magnetism, Institute of Physics, University of Augsburg, D-86135 Augsburg, Germany}

\author{Markus Heyl}
\affiliation{Theoretical Physics III, Center for Electronic Correlations and Magnetism, Institute of Physics, University of Augsburg, D-86135 Augsburg, Germany}
\affiliation{Centre for Advanced Analytics and Predictive Sciences (CAAPS), University of Augsburg, Universitätsstr. 12a, 86159 Augsburg, Germany}

\author{M. A. Rajabpour}
\affiliation{Instituto de F\'isica, Universidade Federal Fluminense, Av.~Gal.~Milton Tavares de Souza s/n, Gragoat\'a, 24210-346, Niter\'oi, RJ, Brazil}
\affiliation{Theoretical Physics III, Center for Electronic Correlations and Magnetism, Institute of Physics, University of Augsburg, D-86135 Augsburg, Germany}

\begin{abstract}
Universal behavior at continuous phase transitions is encoded in long-wavelength fixed-point theories controlled by dimensionality and symmetries. It is a central question for which systems this universal behavior and the underlying long-wavelength description can be captured by a noninteracting theory allowing for analytical and accurate predictions. In this work we provide a systematic procedure for obtaining the best noninteracting fermionic description of a quantum phase transition through fermionic Gaussian states. We demonstrate the performance of our procedure for the interacting long-range transverse-field Ising chain. This formulation reduces the problem to a self-consistent Euler equation that yields a critical line in close agreement with independent numerical benchmarks, while its singular momentum structure provides analytic predictions for continuously varying critical properties. Yet exact diagonalization shows that Gaussian-state fidelity to the exact ground state is poorest, and non-Gaussianity strongest, near the transition. Thus, accurate predictions for critical properties can emerge from the Gaussian variational description even where its microscopic wave-function approximation is weakest.
\end{abstract}

\maketitle

\paragraph{Introduction.}
Fermionic Gaussian states form a low-complexity sector of quantum many-body
Hilbert space: they are efficiently specified by covariance matrices and
underpin protocols for fermionic learning, simulation, and matchgate quantum
information
\cite{Zhao2021,Oszmaniec2022,Mele2025,Ramkumar2026,Helsen2022}.
Recent work has quantified fermionic non-Gaussianity as a many-body and
computational resource \cite{Sierant2026,Tarabunga2026}, while developing
rigorous tests of closeness to the Gaussian manifold and revealing nontrivial
structure even within free-fermionic states
\cite{BittelTrace2025,Collura2026,BittelPAC2025}.
At the same time, free and Gaussian fermions can exhibit universal
entanglement phenomena, thermalization, long-range propagation, topology,
and critical resource signatures
\cite{vanLuijk2025,Bejan2025,Gong2023,Bianchi2021,RamirezRajabpour2026}.
For interacting systems, Gaussian states have long served as variational
manifolds
\cite{RingSchuck1980,BlaizotRipka1986,KrausCirac2010,GuaitaEtAl2019,ShiHubbard2023,Kaicher2023}
and are increasingly used as starting points for more general constructions
\cite{BoutinBauer2021,NiuHolographic2022,LiGfPEPS2023,NiuChiral2024,WuDisentangle2025}.
Related measurement and decomposition schemes use Gaussian structure to
access interacting fermionic observables
\cite{Wan2023,WuKoh2024,Heyraud2025,Majsak2025,WangXu2025}, while systematic
extensions provide controlled routes beyond Gaussianity
\cite{ShiDemlerCirac2018,GuaitaHacklShiDemlerCirac2021}.
These developments raise a sharper question than whether Gaussian states
yield accurate energies: \emph{must a variational manifold faithfully
approximate an interacting critical wave function in order to capture its
universal long-distance physics?}

The ferromagnetic long-range transverse-field Ising chain,
provides a stringent setting for this question. The nearest-neighbor limit is exactly
free-fermionic \cite{LiebSchultzMattis1961,Pfeuty1970}, whereas algebraically
decaying spin couplings generate nonlocal Jordan--Wigner strings and an
interacting fermionic problem. Its ferromagnetic critical behavior spans
long-range mean-field scaling, an interacting regime with continuously
varying exponents, and eventual recovery of short-range Ising criticality;
this structure has been studied by field theory, thermodynamic
linked-cluster expansions, tensor-network methods, and quantum Monte Carlo
\cite{DefenuRMP2023,DuttaBhattacharjee2001,Maghrebi2016,FeySchmidt2016,Jaschke2017,Zhu2018,Puebla2019,Koziol2021,ShirataniTodo2024}.
Entanglement properties of long-range Ising chains have also been studied
\cite{KoffelLewensteinTagliacozzo2012}. Long-range Ising interactions are also directly accessible in
programmable trapped-ion quantum simulators
\cite{PorrasCirac2004,IslamEtAl2013,Monroe2021,LiWuMeiEtAl2023}.
More broadly, Rydberg-atom arrays
\cite{BrowaeysLahaye2020} and ultracold polar molecules
\cite{GorshkovEtAl2011} provide platforms for engineering
long-range interacting spin models.

The precise long-range--to--short-range crossover remains unsettled despite
early renormalization-group and large-$N$ analyses
\cite{FisherMaNickel1972,Sak1973,Suzuki1973,vanEnter1982}, subsequent numerical
and field-theoretic studies supporting different scenarios
\cite{LuijtenBlote2002,Picco2012,BlanchardPiccoRajabpour2013,AngeliniParisiRicciTersenghi2014,BrezinParisiRicciTersenghi2014,PaulosRychkovVanReesZan2016,HoritaSuwaTodo2017,BehanRastelliRychkovZan2017},
and recent large-scale studies of several classical long-range models
together with a quantum $O(n)$ renormalization-group analysis
\cite{XiaoXY2025,LiuPercolation2025,Defenu2017,YaoHeisenberg2025,XiaoIsing2025,LiQuantumRG2026}.
Dangerously irrelevant variables and slow corrections to scaling can
complicate finite-size analyses in parts of the phase diagram
\cite{LangheldKoziolAdelhardtKapferSchmidt2022,Koziol2021,ShirataniTodo2024},
making a direct thermodynamic formulation especially useful. Previous
fermionic-Gaussian work showed that generalized Hartree--Fock states give
accurate results for the long-range transverse-field Ising model in the
antiferromagnetic regime \cite{Kaicher2023}.

Here we develop a thermodynamic Gaussian variational description of the
ferromagnetic long-range transverse-field Ising chain and use it to
investigate how accurately a noninteracting variational manifold can
capture interacting critical behavior. We retain the full spin Hamiltonian
and restrict only the variational state to the fermionic Gaussian manifold.
Translation invariance and Gaussian purity reduce the variational problem
to a momentum-space phase, while the long-range spin interactions remain
encoded nonlinearly through determinant and Pfaffian correlation functions.
The resulting self-consistent Euler equation is the central object of our
analysis. Criticality appears directly as a zero of its Euler field in the thermodynamic limit, yielding an analytic self-consistent condition whose numerical solution closely agrees with independent benchmarks.
\cite{FeySchmidt2016,Jaschke2017,Zhu2018,Koziol2021,ShirataniTodo2024}. The nonanalytic momentum
structure of the same equation, combined with the asymptotic assumptions
specified below, provides closed-form analytic expressions for continuously varying
critical properties.

This agreement coexists with a reduction in microscopic wave-function
accuracy near the transition. Exact diagonalization shows that the
optimized Gaussian state is least faithful to the exact ground state in
the critical region, where the exact state
also exhibits enhanced non-Gaussianity. Thus, the accuracy of the critical
predictions does not track the microscopic state fidelity: the optimized
Gaussian description can retain the long-wavelength structure governing
critical behavior even where its wave-function approximation is weakest.
These results identify the singular structure of the variational solution
as the key to understanding how a Gaussian manifold can capture critical
properties of an interacting system.

\begin{figure}[t]
\centering
\safeincludegraphics[width=\linewidth]{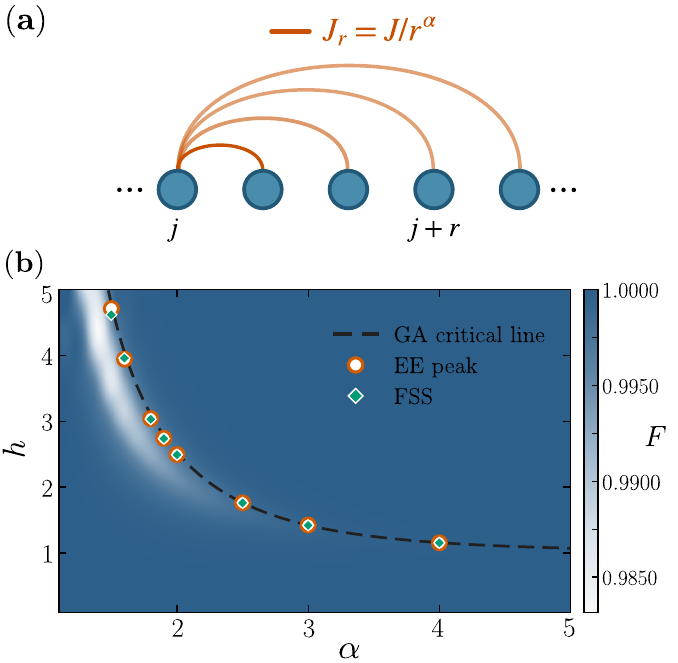}
\caption{\textbf{Gaussian variational description of the long-range
Ising chain.}
\textbf{(a)} Schematic of the chain with ferromagnetic couplings
$J_r=J/r^\alpha$. Representative bonds from one site are shown;
all pairs interact.
\textbf{(b)} Fidelity between the optimized Gaussian state and the
exact finite-size ground state for $L=20$, together with the
thermodynamic Gaussian critical line (dashed curve) and
finite-size estimates of the critical field from the
entanglement-entropy peak (EE peak) and Binder-cumulant
crossings (FSS). The fidelity is reduced most strongly near
the critical line, yet the Gaussian prediction for the
critical field remains in close agreement with the
finite-size estimates. Fields are expressed in units of $J$.}
\label{fig:accuracy-main}
\end{figure}

\paragraph{Model and variational formulation.}
We consider the ferromagnetic long-range transverse-field Ising chain
\begin{equation}
H
=
-\sum_j\sum_{r\ge1}
\frac{J}{r^\alpha}\,
\sigma_j^x\sigma_{j+r}^x
-h\sum_j\sigma_j^z,
\qquad
\alpha>1 ,
\label{eq:H-main}
\end{equation}
where $J>0$ sets the interaction scale, $h$ is the transverse field, and
$\alpha$ controls the interaction range. The long-range interactions are illustrated in Fig.~\ref{fig:accuracy-main}(a). For the finite periodic systems
used in the numerical calculations, the power law is periodized by summing
over all lattice images, so that its fixed-distance thermodynamic limit is
$J_r=J/r^\alpha$; the explicit finite-size coupling is given in
SM Sec.~\ref{sm:finite-size-numerics}.

The nearest-neighbor limit of Eq.~\eqref{eq:H-main} maps under the
Jordan--Wigner transformation to a quadratic fermionic Hamiltonian with
an exactly Gaussian ground state, whereas the long-range spin interaction
generates nonlocal fermionic strings. Our strategy is not to discard these
strings or replace the Hamiltonian by a quadratic one. Instead, we retain
the full spin Hamiltonian and restrict only the variational state to a
fermionic Gaussian manifold. The nonquadratic character of the spin
problem then survives through determinant and Pfaffian expressions for its
correlation functions. The finite-size Gaussian construction, our
Jordan--Wigner conventions, and the corresponding Pfaffian formulas are
given in SM Secs.~\ref{sm:full-gaussian} and \ref{sm:jw-route}.

The apparent many-parameter Gaussian problem simplifies strongly in the
thermodynamic limit. Using the Jordan--Wigner fermions defined in
SM Sec.~\ref{sm:jw-route}, translation invariance makes the mixed Majorana
contraction
\begin{equation}
G_{jk}
=
\left\langle
(c_j-c_j^\dagger)(c_k+c_k^\dagger)
\right\rangle
\label{eq:Gjk-main}
\end{equation}
a Toeplitz matrix specified by a momentum-space symbol $G(q)$. Purity of
the Gaussian state constrains this symbol to unit modulus, so that the
variational state can be parametrized by a single real phase $\varphi_q$,
\begin{equation}
G(q)=-\ee^{\ii\varphi_q},
\qquad
\varphi_{-q}=-\varphi_q .
\label{eq:phase-main}
\end{equation}
This reduction to one function of momentum is central to the analytic
treatment below: the many-body optimization problem becomes a variational
problem for the phase profile $\varphi_q$, while the interacting spin
structure remains encoded in its real-space determinants.

The Fourier coefficients of the symbol are
\begin{equation}
g_n[\varphi]
=
-\frac{1}{\pi}\int_0^\pi dq\,
\cos(\varphi_q-nq).
\label{eq:gn-main}
\end{equation}
They directly determine the spin observables entering the energy. In the
convention used here, $\langle\sigma_j^z\rangle=-g_0$, whereas the
order-parameter correlator takes the Toeplitz-determinant form
\begin{equation}
X_r
\equiv
\langle\sigma_j^x\sigma_{j+r}^x\rangle
=
(-1)^r\det\mathbf T_r,
\qquad
[\mathbf T_r]_{ab}=g_{b-a+1}.
\label{eq:observables-main}
\end{equation}
Thus even though the trial state is Gaussian, the long-range interaction
enters through determinants whose size grows with the spin separation
$r$. This is the source of the nonlinear structure of the resulting
Gaussian variational theory.

Combining the transverse magnetization and all magnetic correlators, the
thermodynamic variational energy density becomes
\begin{equation}
e[\varphi]
=
h\,g_0[\varphi]
-
J\sum_{r\ge1}
\frac{X_r[\varphi]}{r^\alpha}.
\label{eq:energy-main}
\end{equation}
All dependence on the variational state is therefore carried by the single
phase $\varphi_q$, but through the nonlinear Toeplitz determinants
$X_r[\varphi]$. Minimizing this functional leads to the self-consistent
Euler equation that forms the starting point of our critical analysis.
\paragraph{Variational Euler equation.}
The energy functional~\eqref{eq:energy-main} is nonlinear in
$\varphi_q$ through the Toeplitz determinants $X_r[\varphi]$.
At a variational minimum, $\delta e/\delta\varphi_q=0$ for every
momentum, which can be written as
$A_q\sin\varphi_q-B_q\cos\varphi_q=0$, with
\begin{align}
A_q[\varphi,h]
&=
h-\sum_{r\ge1}J_r(-1)^r
\sum_{a,b=1}^{r}
\Delta_{ab}^{(r)}
\cos[q(b-a+1)],
\label{eq:Aq-main}\\
B_q[\varphi]
&=
-\sum_{r\ge1}J_r(-1)^r
\sum_{a,b=1}^{r}
\Delta_{ab}^{(r)}
\sin[q(b-a+1)],
\label{eq:Bq-main}
\end{align}
where
$\Delta_{ab}^{(r)}
=\partial\det\mathbf T_r/\partial[\mathbf T_r]_{ab}$
is the corresponding Toeplitz cofactor. It is useful to combine these two quantities into the complex Euler field
\begin{equation}
\mathcal F_E(q)
\equiv
A_q+\ii B_q .
\label{eq:Euler-field-main}
\end{equation}
The stationarity condition aligns the phase of the Gaussian symbol with
this field.  Selecting the branch that minimizes the variational energy
gives the self-consistency equation
\begin{equation}
G(q)
=
-\frac{\mathcal F_E(q)}{|\mathcal F_E(q)|}
=
-\frac{A_q+\ii B_q}
{\sqrt{A_q^2+B_q^2}} .
\label{eq:Gself-main}
\end{equation}
Equations~\eqref{eq:Aq-main}--\eqref{eq:Gself-main} are exact within the
translation-invariant Gaussian manifold.  Their functional derivation is
given in the End Matter.  Criticality occurs when the Euler field itself
vanishes at the symmetry momentum, which provides the starting point for
the thermodynamic critical-line condition below.

\paragraph{Thermodynamic critical line.}
Along the ferromagnetic branch continuously connected to the
nearest-neighbor Ising critical point, the critical zero of the Euler field
occurs at the Brillouin-zone edge $q=\pi$. At this symmetry momentum
$B_\pi=0$ identically, so the criticality condition
$\mathcal F_E(\pi)=0$ reduces to $A_\pi=0$.

Evaluating the cofactor kernel at $q=\pi$ introduces the alternating
projection
\begin{equation}
K_r[G]
=
(-1)^r
\sum_{a,b=1}^{r}
\Delta_{ab}^{(r)}(-1)^{b-a+1},
\label{eq:Kr-main}
\end{equation}
where the cofactors $\Delta_{ab}^{(r)}$ are determined by the same
Gaussian symbol $G$. The condition $A_\pi=0$ then gives
\begin{equation}
h_c
=
J\sum_{r=1}^{\infty}
\frac{K_r[G_c]}{r^\alpha},
\label{eq:hc-main}
\end{equation}
with $G_c$ the self-consistent critical symbol.

Equations~\eqref{eq:Gself-main} and \eqref{eq:hc-main} therefore form a
coupled nonlinear thermodynamic problem for $G_c(q)$ and $h_c$.
The transition is obtained directly from the variational Euler equation,
without invoking finite-size crossings. 

We solve the coupled equations numerically in momentum space and extrapolate
 the interaction-range cutoff $r_{\max}\to\infty$; implementation details
 are given in the End Matter. Exact-diagonalization results for the
 fidelity at $L=20$, shown in Fig.~\ref{fig:accuracy-main}(b), reveal that the Gaussian approximation is
 least accurate near the critical line, where the exact ground state also
 shows its strongest departure from Gaussianity (see SM).
 Nevertheless, the resulting thermodynamic critical line, also shown in
 Fig.~\ref{fig:accuracy-main}(b), agrees closely with independent finite-size
 estimates from the entanglement-entropy peak and Binder-cumulant crossings.
 A comparison with quantum Monte Carlo results~\cite{ShirataniTodo2024},
 presented in the End Matter [Fig.~\ref{fig:em-critical-line-qmc}(a)],
 shows similarly close agreement.
 Further diagnostics and finite-size analyses are given in
 SM Sec.~\ref{sm:finite-size-numerics}.

\paragraph{Analytic magnetic correlation exponent.}
The same critical Euler field that determines $h_c$ also controls the
long-distance magnetic correlations. At the self-consistent critical
solution, define its magnitude
$\mathcal E_E(q)=|\mathcal F_E(q)|$ and characterize the zero at
$q=\pi$ by
\begin{equation}
\mathcal E_E(\pi-k)\sim |k|^{\zeta_E},
\qquad k\to0 .
\label{eq:zeta-def-main}
\end{equation}
Here $\zeta_E$ is a static momentum-space exponent of the variational
stationarity equation. We seek the magnetic
correlation exponent defined by
$X_r=\langle\sigma_j^x\sigma_{j+r}^x\rangle\sim r^{-x_\sigma}$.

\begin{figure}[t]
\centering

\safeincludegraphics[width=\columnwidth]
{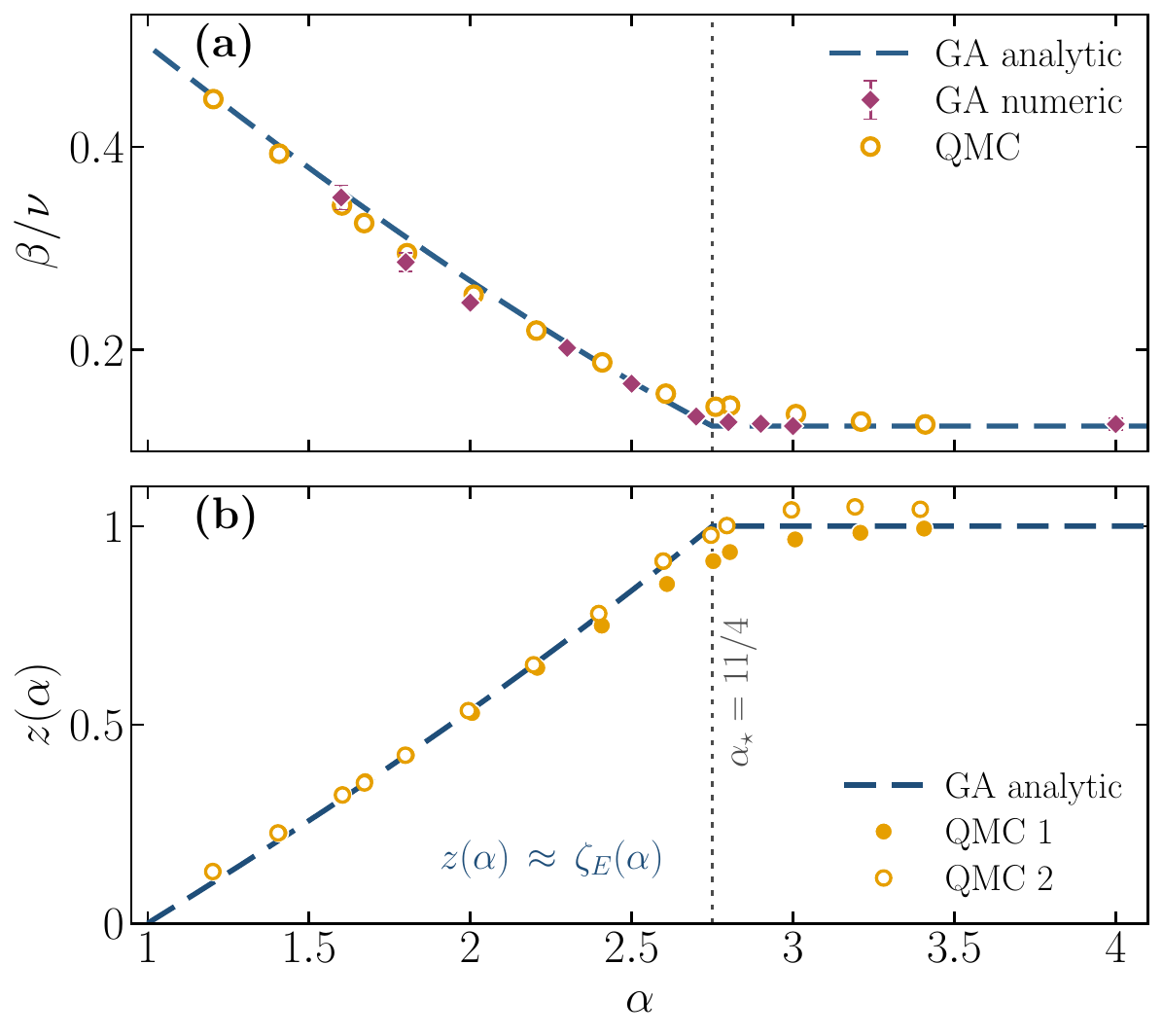}

\caption{
\textbf{Critical exponents.}
\textbf{(a)} Magnetic exponent $\beta/\nu$, comparing Gaussian finite-size results, benchmark data, and the analytic prediction $\beta/\nu=x_\sigma/2$.
\textbf{(b)} Dynamical exponent $z$: QMC estimates from
Ref.~\cite{ShirataniTodo2024} are compared with the static Euler exponent
$\zeta_E(\alpha)$, providing a numerical test of the predicted relation
$z=\zeta_E$.
}
\label{fig:hc_exponent_main}
\end{figure}

The analytic extraction of $x_\sigma$ uses three asymptotic ingredients,
discussed in detail in SM Sec.~\ref{sm:analytic-closures}. First, the
alternating projection of the inverse Toeplitz matrix grows linearly with
$r$ at criticality. This behavior is exact at the short-range critical
point and for the canonical single-jump Fisher--Hartwig symbol, and we use
its continuation along the optimized critical branch. Together with the
exact cofactor identity, this gives
\begin{equation}
K_r\sim \mathcal D_K\,r^{1-x_\sigma}.
\label{eq:Kr-scaling-main}
\end{equation}
Second, the momentum-resolved cofactor kernel has the scaling form
$\mathcal K_r(k)/K_r\to\Psi(kr)$ in the joint limit
$r\to\infty$, $k\to0$ with $kr$ fixed. The canonical single-jump problem
admits an exact Cauchy-matrix representation that supports this scaling
and fixes a nonzero linear small-$kr$ term; the corresponding derivation
and short-range checks are given in SM Sec.~\ref{sm:analytic-closures}.

To insert this real-space scaling back into the Euler equation, let
$\mathcal K_r(k)$ denote the momentum-resolved version of $K_r$, with
$\mathcal K_r(0)=K_r$. Using the critical condition
$\mathcal F_E(\pi)=0$, the field near the Brillouin-zone edge can be written
as
\begin{equation}
\mathcal F_E(\pi-k)
=
J\sum_{r\ge1}
\frac{K_r-\mathcal K_r(k)}{r^\alpha}.
\label{eq:Euler-subtracted-main}
\end{equation}
This form separates the critical cancellation from the momentum-dependent
correction.

There are two contributions to its small-$k$ behavior. The large-$r$
tail of Eq.~\eqref{eq:Euler-subtracted-main}, together with
Eq.~\eqref{eq:Kr-scaling-main} and the scaling form above, produces a
nonanalytic term, while the regular fixed-$r$ part admits an ordinary
Taylor expansion:
\begin{equation}
\begin{aligned}
\mathcal F_E(\pi-k)
&=
\mathcal F_E^{\rm reg}(\pi-k)
+
\mathcal F_E^{\rm LR}(\pi-k),\\
\mathcal F_E^{\rm reg}(\pi-k)
&=O(k),\\
\bigl|\mathcal F_E^{\rm LR}(\pi-k)\bigr|
&\sim |k|^{\alpha+x_\sigma-2}.
\end{aligned}
\label{eq:Euler-two-parts-main}
\end{equation}
The Mellin analysis leading to the last line, including the conditions for
the convergence of the long-range contribution, is given in
SM Sec.~\ref{sm:analytic-closures}. Provided the corresponding amplitudes
are nonzero and do not cancel, the smaller power controls the magnitude of
the Euler field, so that
\begin{equation}
\zeta_E
=
\min\{\alpha+x_\sigma-2,\,1\}.
\label{eq:zeta-min-main}
\end{equation}
Thus the continuously varying branch is controlled by the nonanalytic
large-distance tail, whereas the short-range branch is controlled by the
regular linear term.

The third ingredient concerns the phase rather than the magnitude of the
Euler field. Because of the stationarity condition~\eqref{eq:Gself-main},
normalization removes the magnitude of $\mathcal F_E(q)$ and transfers its
phase directly to the critical Gaussian symbol $G_c(q)$. A zero of
$\mathcal F_E$ can therefore generate a singular phase structure in the
Toeplitz symbol controlling $X_r$. The magnitude exponent $\zeta_E$ alone
does not determine this phase jump. We close the asymptotic problem by
following the single-jump Fisher--Hartwig branch continuously connected to
the short-range Ising solution. In this branch the shifted Toeplitz symbol
fixes the Fisher--Hartwig exponent once the endpoint phase is known, while
the remaining phase-locking condition relates that endpoint phase to the
Euler singularity. This gives
\begin{equation}
\delta=1-\frac{\zeta_E}{2},
\qquad
x_\sigma=\delta^2 .
\label{eq:phase-FH-main}
\end{equation}
The first relation is therefore the phase closure of the Gaussian
asymptotic theory, while the second is the single-jump Fisher--Hartwig
exponent of $X_r=(-1)^r\det\mathbf T_r$. Exact short-range and canonical
single-jump checks, together with the precise residual phase assumption,
are given in SM Sec.~\ref{sm:analytic-closures}.

Combining Eqs.~\eqref{eq:zeta-min-main} and
\eqref{eq:phase-FH-main} therefore closes the problem and gives
\begin{equation}
x_\sigma^{GA}(\alpha)
=
\begin{cases}
\left(\sqrt{5-\alpha}-1\right)^2,
&1<\alpha<11/4,\\[1mm]
1/4,
&\alpha>11/4.
\end{cases}
\label{eq:xsigma-main}
\end{equation}
The crossover occurs when the nonanalytic long-range power reaches the
regular linear contribution, yielding $\alpha_\star=11/4$. At this point
the two powers coincide and logarithmic corrections are not excluded.

The distinction between exact Gaussian identities, the asymptotic
closures entering the analytic solution, and their interpretation in
terms of the physical long-range fixed point is discussed in
SM Sec.~\ref{sm:status-critical-relations}.

The static Euler exponent can also be related to the physical dynamical
exponent. At a critical point governed by the long-range interaction,
scale invariance gives
\begin{equation}
z=\alpha+x_\sigma-2 .
\label{eq:z-scaling-main}
\end{equation}
Independently, on the long-range branch
Eq.~\eqref{eq:zeta-min-main} gives
$\zeta_E=\alpha+x_\sigma-2$, and therefore $z=\zeta_E$.
This identification does not require interpreting the Euler field as an
excitation spectrum: the two exponents are governed by the same
long-range scaling combination. In the short-range Ising regime,
$z=1$ and the regular Euler contribution independently gives
$\zeta_E=1$. The scaling argument is detailed in the End Matter.

Figure~\ref{fig:hc_exponent_main}(a) compares the analytic Gaussian
prediction for $x_\sigma$ with finite-size estimates from the optimized spin
states and with available numerical benchmarks. Where ordinary finite-size
scaling applies, $x_\sigma=2\beta/\nu$, and the analytic result reproduces
both the continuously varying long-range regime and the crossover to the
short-range Ising value $\beta/\nu=1/8$.
Figure~\ref{fig:hc_exponent_main}(b) compares the resulting prediction
$z(\alpha)=\zeta_E(\alpha)$ with numerical estimates of the dynamical
exponent obtained from QMC, showing close agreement across the range of
$\alpha$ considered.

\begin{figure}[t]
\centering
\safeincludegraphics[width=\columnwidth]{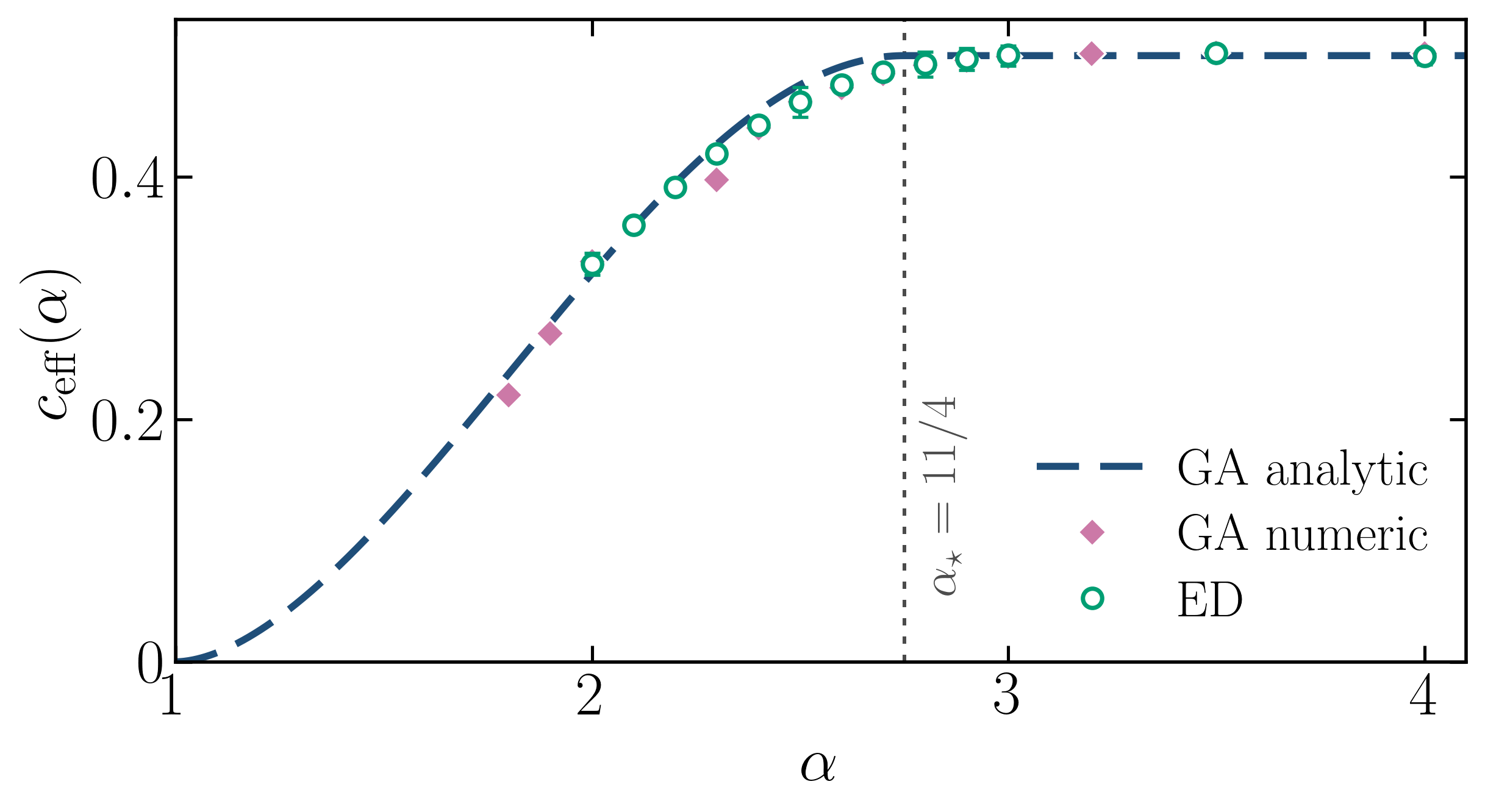}
\caption{\textbf{Logarithmic entanglement coefficient.}
Numerical estimates of $c_{\rm eff}$ from optimized Gaussian
states and ED, compared with the analytic dilogarithm expression
Eq.~\eqref{eq:ceff-main}. The short-range branch approaches
$c_{\rm eff}=1/2$. In the long-range regime the quantity is interpreted as
a logarithmic entanglement coefficient rather than automatically as a
conformal central charge.}
\label{fig:ceff-main}
\end{figure}

\paragraph{Entanglement and logarithmic coefficient.}
The critical Gaussian symbol also determines the entanglement properties
of the variational state.  For a spatial block, the von Neumann entropy is
obtained exactly from the eigenvalues of its restricted Majorana covariance
matrix using the standard Gaussian-state construction
\cite{Peschel2003,PeschelEisler2009}; the explicit formula used in our
numerics is given in SM Sec.~\ref{sm:finite-size-numerics}.  At criticality we
characterize its leading growth by
\begin{equation}
S_1(\ell)
=
\frac{c_{\rm eff}(\alpha)}{3}\log\ell+B+o(1).
\label{eq:ceff-def-main}
\end{equation}
In the genuinely long-range regime, $c_{\rm eff}$ parametrizes the
logarithmic entanglement growth and is not assumed to be the central
charge of an underlying conformal field theory.

To determine this coefficient analytically, one additional spectral
identification is required. We assume that, after removal of smooth
factors irrelevant to the leading logarithm, the restricted covariance
operator belongs to the same asymptotic spectral class as the corresponding
single-jump critical fermionic problem. The relation between singular fermionic correlation symbols, Toeplitz
asymptotics, and logarithmic entanglement scaling is well established
\cite{JinKorepin2004,ItsJinKorepin2005,AresEtAl2015,AresEtAl2018}.
For the corresponding exactly solvable free-fermion interface problem,
the entanglement spectrum and logarithmic coefficient are known exactly
\cite{EislerPeschel2010,PeschelEisler2012}.
We make the additional spectral identification
$s=\sin(\pi\delta)$, with $\delta=1-\zeta_E/2$, which gives
\begin{equation}
\begin{multlined}
c_{\rm eff}^{GA}
=-\frac{3}{\pi^2}\bigl\{
(1+s)\operatorname{Li}_2(-s)
+(1-s)\operatorname{Li}_2(s)
\\
+\log s\bigl[
(1-s)\log(1-s)
+(1+s)\log(1+s)
\bigr]\bigr\}.
\end{multlined}
\label{eq:ceff-main}
\end{equation}
where $\operatorname{Li}_2$ is the dilogarithm.  On the long-range branch,
$\delta=\sqrt{5-\alpha}-1$, so that
$s=\sin[\pi(\sqrt{5-\alpha}-1)]$ for $1<\alpha<11/4$.
For $\alpha>11/4$, $\delta=1/2$ and hence $s=1$, reducing
Eq.~\eqref{eq:ceff-main} to $c_{\rm eff}^{GA}=1/2$.

The covariance-matrix entropy itself is exact within the Gaussian
variational state, whereas the thermodynamic expression
Eq.~\eqref{eq:ceff-main} relies on the additional spectral identification
above.  Its derivation and precise logical status are given in the End
Matter.  The independent numerical extraction of $c_{\rm eff}$ from
block and half-chain entropies is described in
SM Sec.~\ref{sm:finite-size-numerics}.  Figure~\ref{fig:ceff-main} compares these finite-size estimates with the analytic Gaussian prediction, showing good agreement with the exact-diagonalization (ED) results.

\paragraph{Discussion.}
Our results show that universal critical information can survive a severe
restriction of the variational manifold even when the wave-function
description deteriorates near criticality. The self-consistent Euler equation determines the thermodynamic critical line, while its Brillouin-zone-edge singularity together with the asymptotic Toeplitz structure determines the magnetic exponent. Combined with long-range critical scaling, the same endpoint exponent gives the dynamical relation $z=\zeta_E$, and, with an additional spectral closure, the logarithmic entanglement coefficient. Exact diagonalization makes the contrast particularly striking: the exact ground state is most strongly non-Gaussian and the optimized Gaussian state least faithful near criticality (see SM), precisely where the critical properties remain accurately reproduced. The relevant information is therefore not encoded in microscopic wave-function fidelity, but in the singular structure retained by the optimized Gaussian solution. More broadly, our results suggest that the reach of a variational theory of criticality is set by two ingredients: whether the manifold can represent the infrared structure controlling a universal quantity, and whether variational optimization selects that structure. Gaussian manifolds may therefore remain effective not only at Gaussian or mean-field fixed points, but also at interacting critical points whose essential infrared data can be encoded through mass zeros, soft modes, or singular covariance structure. Their limitations should instead arise when the universal physics depends on additional infrared parameters, such as a Luttinger parameter or compactification radius, that are not generically encoded by the Gaussian covariance structure. The relevant question is thus not whether the critical state is itself close to Gaussian, but whether the infrared information controlling the observable of interest is representable within the variational manifold.

\par
\section*{Acknowledgments }
\noindent 
M.A.R. acknowledges partial support from CNPq and FAPERJ (grant number E-26/210.062/2023). We gratefully acknowledge the resources on the LiCCA HPC cluster of the University of Augsburg, co-funded by the Deutsche Forschungsgemeinschaft (DFG, German Research Foundation)–Project-ID 499211671. This work was supported by the research funding program “Forschungspotenziale besser nutzen!” of the University of Augsburg. 

\bibliographystyle{apsrev4-2-italic-force-titles}
\bibliography{references}


\section*{End Matter}
The thermodynamic Euler equation and the critical-line condition are exact
within the translation-invariant Gaussian variational manifold.  Building
on these exact relations, the magnetic exponent and entanglement
coefficient require additional asymptotic closures, specified below.  We
collect the corresponding analytic steps here, while the numerical
implementation and finite-size analyses are given in the Supplemental
Material.

\paragraph{Thermodynamic critical line.}
The variational energy depends on the phase $\varphi_q$ only through the
Fourier coefficients $g_n$ and the Toeplitz determinants built from them.
From Eq.~\eqref{eq:gn-main},
\begin{align}
\delta g_n
&=
\frac{1}{\pi}\int_0^\pi dq\,
\sin(\varphi_q-nq)\,\delta\varphi_q ,
\label{eq:em-dgn}
\\
\delta X_r
&=
(-1)^r\sum_{a,b=1}^{r}
\Delta_{ab}^{(r)}\,\delta g_{b-a+1},
\label{eq:em-dX}
\end{align}
where
$\Delta_{ab}^{(r)}
=\partial\det\mathbf T_r/\partial[\mathbf T_r]_{ab}$.
Thus the variation of each long-range spin correlator is reduced to the
variation of the same phase profile that parametrizes the Gaussian state.

Substitution into $\delta e[\varphi]=0$ gives
\begin{equation}
h\sin\varphi_q
=
\sum_{r\ge1}J_r(-1)^r
\sum_{a,b=1}^{r}
\Delta_{ab}^{(r)}
\sin\!\left[\varphi_q-q(b-a+1)\right].
\label{eq:em-rawEuler}
\end{equation}
Although this equation has the form of a momentum-space alignment
condition, it remains nonlinear because the cofactors
$\Delta_{ab}^{(r)}$ are themselves functionals of the symbol $G$.
Separating the coefficients of $\sin\varphi_q$ and
$\cos\varphi_q$ reproduces Eqs.~\eqref{eq:Aq-main} and
\eqref{eq:Bq-main}.  Choosing the minimizing stationary branch then gives
the self-consistency equation~\eqref{eq:Gself-main}.

\begin{figure}[t]
    \centering
    \includegraphics[width=\columnwidth]{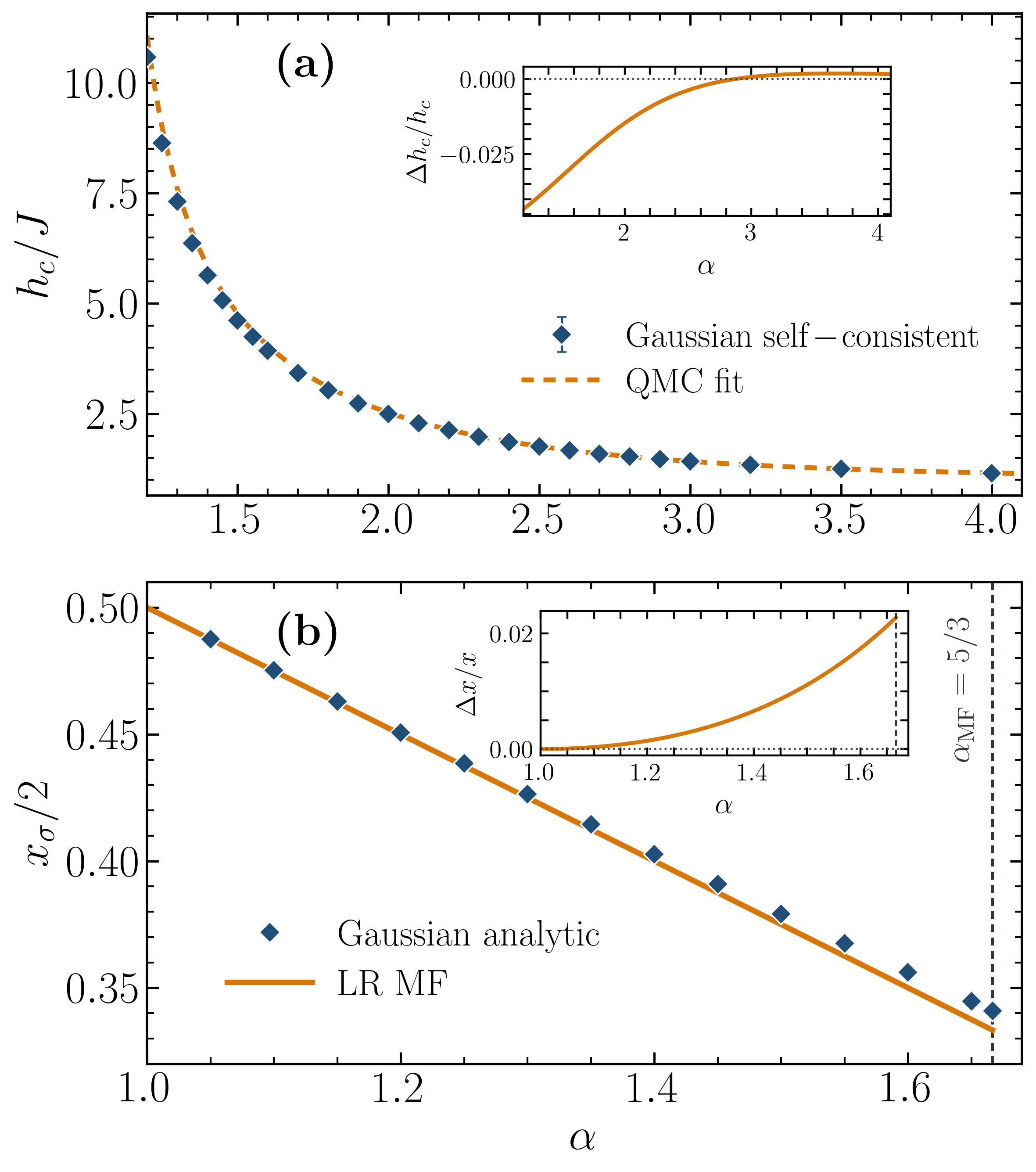}
   \caption{\textbf{Gaussian critical line and magnetic exponent.}
(a) Blue diamonds show the thermodynamic self-consistent Gaussian critical
fields after extrapolation $r_{\max}\to\infty$; the orange dashed line is the
QMC fit~\cite{ShirataniTodo2024}. The inset shows their signed relative
deviation.
(b) Blue diamonds show the analytic Gaussian prediction for $x_\sigma/2$,
while the orange line is the LR MF result $(3-\alpha)/4$. The dashed vertical
line marks $\alpha_{\rm MF}=5/3$; the inset shows the relative difference
between the two predictions.}
    \label{fig:em-critical-line-qmc}
\end{figure}

Along the ferromagnetic critical branch, the zero of the Euler field occurs
at the Brillouin-zone edge $q=\pi$.  At this momentum the sine contribution
vanishes identically,
$\sin[\pi(b-a+1)]=0$, and therefore $B_\pi=0$.  The criticality condition
is consequently reduced to the single real equation $A_\pi=0$, with
\begin{equation}
A_\pi
=
h-J\sum_{r\ge1}\frac{K_r[G]}{r^\alpha},
\label{eq:em-Api}
\end{equation}
where $K_r[G]$ is the alternating cofactor projection defined in
Eq.~\eqref{eq:Kr-main}.  Hence $\mathcal F_E(\pi)=0$ gives directly
\begin{equation}
h_c
=
J\sum_{r\ge1}\frac{K_r[G_c]}{r^\alpha}.
\label{eq:em-hc}
\end{equation}
The condition is scalar at $q=\pi$, but it remains self-consistent:
the coefficients $K_r[G_c]$ are evaluated from the same critical symbol
$G_c$ obtained from Eq.~\eqref{eq:Gself-main}.  The pair
$(G_c,h_c)$ is therefore determined simultaneously in the thermodynamic
limit. As an independent benchmark of the thermodynamic critical line, we compare the self-consistent Gaussian solution with the QMC
parametrization~\cite{ShirataniTodo2024} in
Fig.~\ref{fig:em-critical-line-qmc}(a). The agreement is close over the full
range shown, with the largest deviation occurring in the strongly
long-range regime. The Gaussian points are obtained directly from the coupled
Euler equation and criticality condition.

\paragraph{Brillouin-zone-edge scaling and magnetic exponent.}
We now justify the asymptotic steps connecting the critical Euler field to
the magnetic exponent.  The first ingredient concerns the large-$r$
behavior of the critical-field kernel.  Whenever $\mathbf T_r$ is
nonsingular, the cofactor identity gives
\begin{equation}
K_r
=
-X_r\,
\bm\eta^{\mathsf T}\mathbf T_r^{-1}\bm\eta,
\qquad
\eta_a=(-1)^a .
\label{eq:em-Kinverse}
\end{equation}
At the short-range critical point the alternating projection is exactly
$\bm\eta^{\mathsf T}[\mathbf T_r^{(0)}]^{-1}\bm\eta=-\pi r/2$.
Our first asymptotic closure is that its linear growth persists along the
long-range critical branch,
$\bm\eta^{\mathsf T}\mathbf T_r^{-1}\bm\eta\sim-\kappa r$ with
$0<\kappa<\infty$.  Together with
$X_r\sim r^{-x_\sigma}$, this gives the scaling
$K_r\sim r^{1-x_\sigma}$ used in Eq.~\eqref{eq:Kr-scaling-main}.

To determine how this real-space tail enters the Euler field near
$q=\pi$, define
\begin{align}
\mathcal K_r(k)
&=
(-1)^r
\sum_{a,b=1}^{r}
\Delta_{ab}^{(r)}
(-1)^{b-a+1}
\ee^{-\ii k(b-a+1)},
\label{eq:em-Kk}\\
\frac{\mathcal K_r(k)}{K_r}
&\longrightarrow\Psi(kr),
\qquad
\Psi(0)=1 .
\label{eq:em-Psi}
\end{align}
By construction, $\mathcal K_r(0)=K_r$.  The second line is our
asymptotic scaling closure, taken for $r\to\infty$, $k\to0$ with
$kr$ fixed. Inserting this form into the exact
critical subtraction~\eqref{eq:Euler-subtracted-main} gives for the
large-distance contribution
\begin{equation}
\mathcal F_E^{\rm LR}(\pi-k)
\sim
J\mathcal D_K
\sum_{r\ge1}
\frac{1-\Psi(kr)}{r^{1+\mu}},
\qquad
\mu=\alpha+x_\sigma-2 .
\label{eq:em-endpoint-sum}
\end{equation}
For $0<\mu<1$, the change of variable $u=|k|r$ converts the asymptotic
tail into
\begin{equation}
\mathcal F_E^{\rm LR}(\pi-k)
\sim
|k|^\mu
\int_0^\infty
du\,u^{-1-\mu}
\bigl[1-\Psi(\operatorname{sgn}k\,u)\bigr],
\label{eq:em-Mellin}
\end{equation}
up to an overall nonzero amplitude.  This establishes the nonanalytic
power $|\mathcal F_E^{\rm LR}|\sim
|k|^{\alpha+x_\sigma-2}$ quoted in the main text.

The regular contribution has a different origin.  For every fixed $r$,
$\mathcal K_r(k)$ has an ordinary Taylor expansion about $k=0$, whose
leading nonvanishing term is generically linear.  The full Euler field
therefore contains both the analytic $O(k)$ contribution and the
nonanalytic large-$r$ contribution above.  Their competition gives
Eq.~\eqref{eq:zeta-min-main},
$\zeta_E=\min\{\alpha+x_\sigma-2,1\}$.

It remains to connect this magnitude exponent to the magnetic correlator.
By Eq.~\eqref{eq:Gself-main}, normalization removes
$|\mathcal F_E|$ and leaves its phase as the phase of the critical symbol
$G_c$.  Our phase closure assumes that, after removal of smooth
nonvanishing factors, this singular part is represented by a single
Fisher--Hartwig jump continuously connected to the short-range Ising
branch.  Its jump strength satisfies
$\delta=1-\zeta_E/2$, while the corresponding Toeplitz determinant decays
as $r^{-\delta^2}$.  This gives precisely
Eq.~\eqref{eq:phase-FH-main}.

On the long-range branch,
$\zeta_E=\alpha+x_\sigma-2$ together with
$x_\sigma=(1-\zeta_E/2)^2$ gives
$\delta^2+2\delta+\alpha-4=0$.  The physical root,
$\delta=\sqrt{5-\alpha}-1$, yields
Eq.~\eqref{eq:xsigma-main}.
For comparison, Fig.~\ref{fig:em-critical-line-qmc}(b) also shows the
long-range mean-field (LR MF) result
$x_\sigma/2=(3-\alpha)/4$ in its regime of validity,
$1<\alpha\leq5/3$.  The Gaussian prediction closely follows the LR MF
behavior throughout this regime, with their relative difference shown in
the inset.
The long-range power reaches the regular
linear contribution at $\zeta_E=1$, for which
$x_\sigma=1/4$ and hence $\alpha_\star=11/4$.  At this coincidence of
powers logarithmic corrections are not excluded.


\paragraph{Entanglement-spectrum closure.}
The finite-block entropy is obtained exactly from the restricted Gaussian
covariance matrix, as described in the Supplemental Material. To obtain
its logarithmic coefficient analytically, we make the additional spectral
identification stated in the main text: after removal of smooth factors,
the singular part of the critical covariance operator is assumed to lie in
the same asymptotic spectral class as a single-jump fermionic problem
\cite{JinKorepin2004,ItsJinKorepin2005,AresEtAl2015,AresEtAl2018}.
We identify its Fisher--Hartwig jump $\delta$ with the spectral parameter
of the exactly solvable free-fermion interface problem through
$s=\sin(\pi\delta)$.

In the interface problem, the reference spectral parameter $\epsilon$ and
the entanglement-spectrum parameter $\omega$ are related by
\begin{equation}
\cosh\omega(\epsilon)
=
\frac{\cosh\epsilon}{s},
\label{eq:em-omega}
\end{equation}
\cite{EislerPeschel2010,PeschelEisler2012}. A mode characterized by
$\omega$ contributes
$h(\omega)=\log(1+\ee^{-2\omega})
+2\omega/(\ee^{2\omega}+1)$
to the von Neumann entropy. Using the corresponding asymptotic spectral
density gives
\begin{equation}
I(s)
=
\int_0^\infty d\epsilon\,
h[\omega(\epsilon)],
\qquad
c_{\rm eff}
=
\frac{6I(s)}{\pi^2},
\label{eq:em-ceffI}
\end{equation}
following the exact free-fermion result
\cite{EislerPeschel2010,PeschelEisler2012}. The factor of two implicit in
this normalization accounts for the two ends of the interval.

The integral can be evaluated exactly \cite{EislerPeschel2010}.
Differentiating twice with respect to $s$ gives
\begin{equation}
I''(s)
=
-\frac{\log s}{1-s^2},
\qquad
I(0)=I'(0)=0,
\label{eq:em-Ieq}
\end{equation}
whose integration yields
\begin{equation}
\begin{aligned}
I(s)=-\frac12\Big[&
(1+s)\operatorname{Li}_2(-s)
+(1-s)\operatorname{Li}_2(s)
\\
&+(1-s)\log s\,\log(1-s)
\\
&+(1+s)\log s\,\log(1+s)
\Big].
\end{aligned}
\label{eq:em-Idilog}
\end{equation}
Together with Eq.~\eqref{eq:em-ceffI}, this reproduces the analytic
coefficient in Eq.~\eqref{eq:ceff-main}. On the short-range branch
$\delta=1/2$, hence $s=1$ and $I(1)=\pi^2/12$, giving
$c_{\rm eff}=1/2$.

Thus the covariance-matrix entropy is exact within the Gaussian state,
whereas the closed thermodynamic coefficient requires the spectral
identification~\eqref{eq:em-omega}. Once this identification is made, the
remaining spectral integral and its dilogarithm evaluation coincide with
the exact free-fermion result of
Refs.~\cite{EislerPeschel2010,PeschelEisler2012}.

\clearpage
\onecolumngrid
\setcounter{section}{0}
\renewcommand{\thesection}{S\arabic{section}}
\renewcommand{\thesubsection}{\thesection.\Alph{subsection}}
\renewcommand{\thesubsubsection}{\thesubsection.\arabic{subsubsection}}
\renewcommand{\theHsection}{supp.\arabic{section}}
\renewcommand{\theHsubsection}{supp.\arabic{section}.\arabic{subsection}}
\renewcommand{\theHsubsubsection}{supp.\arabic{section}.\arabic{subsection}.\arabic{subsubsection}}

\setcounter{equation}{0}
\renewcommand{\theequation}{S\arabic{equation}}
\renewcommand{\theHequation}{S\arabic{equation}}
\setcounter{figure}{0}
\renewcommand{\thefigure}{S\arabic{figure}}
\renewcommand{\theHfigure}{S\arabic{figure}}
\setcounter{table}{0}
\renewcommand{\thetable}{S\arabic{table}}
\renewcommand{\theHtable}{S\arabic{table}}

\setcounter{section}{0}
\setcounter{subsection}{0}
\setcounter{secnumdepth}{3}
\renewcommand{\thesection}{S\arabic{section}}
\renewcommand{\thesubsection}{\thesection.\Alph{subsection}}
\renewcommand{\thesubsubsection}{\thesubsection.\arabic{subsubsection}}
\renewcommand{\theHsection}{S\arabic{section}}
\renewcommand{\theHsubsection}{S\arabic{section}.\Alph{subsection}}
\renewcommand{\theHsubsubsection}{S\arabic{section}.\Alph{subsection}.\arabic{subsubsection}}

\makeatletter
\def\@seccntformat#1{\csname the#1\endcsname\quad}
\makeatother

\begin{center}
{\large\bf Supplemental Material}
\end{center}

\section{Analytic support for the asymptotic closures}
\label{sm:analytic-closures}

Here we collect the analytic steps underlying the three asymptotic
ingredients used to extract the magnetic correlation exponent in the main
text. It is useful to separate identities that are exact within the
Gaussian manifold from statements that require an asymptotic continuation
along the optimized long-range critical branch. The first closure can be
proved exactly for the short-range critical symbol and, more generally,
for a canonical single-jump Fisher--Hartwig family. The second admits an
explicit Cauchy-matrix representation in the same family and leads directly
to the $kr$ scaling and Mellin power counting. The final relation between
the magnitude exponent of the Euler field and the Fisher--Hartwig jump is
a phase-locking condition; its short-range limit and its winding structure
are exact, while the continuation of the endpoint phase along the
long-range branch is the remaining nontrivial closure.

\subsection{Exact cofactor identity and inverse-Toeplitz scaling}

Recall
\begin{equation}
X_r=(-1)^r\det T_r,
\qquad
[T_r]_{ab}=g_{b-a+1},
\end{equation}
and let
\begin{equation}
\eta_a=(-1)^a .
\end{equation}
Writing the cofactor matrix as
$\Delta_{ab}^{(r)}=\det(T_r)(T_r^{-1})_{ba}$ gives the exact identity
\begin{equation}
K_r
=
-X_r\,\bm\eta^{\mathsf T}T_r^{-1}\bm\eta .
\label{eq:sm-K-cofactor}
\end{equation}
Thus the large-$r$ behavior of $K_r$ is fixed once the alternating
projection of the inverse Toeplitz matrix is known.

At the nearest-neighbor critical point the Gaussian phase is
$\varphi_q^{(0)}=q/2$. Hence
\begin{equation}
g_n^{(0)}
=
-\frac{1}{\pi}\int_0^\pi dq\,
\cos\!\left[\left(\frac12-n\right)q\right]
=
\frac{(-1)^n}{\pi(n-\frac12)} .
\label{eq:sm-gn-SR}
\end{equation}
It follows that
\begin{equation}
[T_r^{(0)}]_{ab}
=
\frac{(-1)^{b-a+1}}
{\pi(b-a+\frac12)} .
\end{equation}
Introduce
\begin{equation}
D=\operatorname{diag}[(-1)^1,\ldots,(-1)^r],
\qquad
[C_{1/2}]_{ab}=\frac{1}{b-a+\frac12},
\end{equation}
so that
\begin{equation}
T_r^{(0)}=-\frac{1}{\pi}D C_{1/2}D,
\qquad
\bm\eta=D\bm 1.
\end{equation}
Therefore
\begin{equation}
\bm\eta^{\mathsf T}[T_r^{(0)}]^{-1}\bm\eta
=
-\pi\,\bm 1^{\mathsf T}C_{1/2}^{-1}\bm 1 .
\label{eq:sm-altproj-C}
\end{equation}

The remaining Cauchy quadratic form can be evaluated exactly. More
generally, define
\begin{equation}
[C_\lambda]_{ab}=\frac{1}{b-a+\lambda},
\qquad 0<\lambda<1,
\end{equation}
and let $\bm x=C_\lambda^{-1}\bm 1$. The equations
$C_\lambda\bm x=\bm 1$ read
\begin{equation}
\sum_{b=1}^r\frac{x_b}{b-a+\lambda}=1,
\qquad a=1,\ldots,r.
\end{equation}
Introduce the rational function
\begin{equation}
F(z)=\sum_{b=1}^r\frac{x_b}{b-z}.
\end{equation}
Since $F(a-\lambda)=1$ for all $a=1,\ldots,r$, its zeros and poles fix it
to
\begin{equation}
F(z)
=
1-
\frac{\displaystyle\prod_{a=1}^r[z-(a-\lambda)]}
{\displaystyle\prod_{b=1}^r(z-b)} .
\label{eq:sm-rational-F}
\end{equation}
At large $z$,
\begin{equation}
F(z)
=
-\frac{r\lambda}{z}+O(z^{-2}),
\end{equation}
whereas its definition gives
\begin{equation}
F(z)
=
-\frac{\sum_bx_b}{z}+O(z^{-2}).
\end{equation}
Hence
\begin{equation}
\bm 1^{\mathsf T}C_\lambda^{-1}\bm 1=r\lambda .
\label{eq:sm-Cauchy-sum}
\end{equation}
For $\lambda=1/2$, Eqs.~\eqref{eq:sm-altproj-C} and
\eqref{eq:sm-Cauchy-sum} give the exact short-range identity
\begin{equation}
\bm\eta^{\mathsf T}[T_r^{(0)}]^{-1}\bm\eta
=
-\frac{\pi r}{2}.
\label{eq:sm-SR-altproj-exact}
\end{equation}

The same calculation extends to the canonical one-jump family
\begin{equation}
G_s(q)=-e^{isq},
\qquad
0<s<1,
\qquad
\lambda=1-s.
\label{eq:sm-canonical-symbol}
\end{equation}
Its Fourier coefficients are
\begin{equation}
g_n^{(s)}
=
\frac{(-1)^n\sin(\pi s)}{\pi(n-s)},
\end{equation}
so that
\begin{equation}
T_r^{(s)}
=
-\frac{\sin(\pi s)}{\pi}\,
D C_\lambda D .
\end{equation}
Using Eq.~\eqref{eq:sm-Cauchy-sum},
\begin{equation}
\bm\eta^{\mathsf T}[T_r^{(s)}]^{-1}\bm\eta
=
-\frac{\pi(1-s)}{\sin(\pi s)}\,r .
\label{eq:sm-canonical-linear}
\end{equation}
Thus the linear growth of the inverse-Toeplitz projection is exact for the
entire canonical single-jump family, not only at the nearest-neighbor
critical point.

For the optimized long-range critical symbol we write the singular part
as a single jump multiplied by a smooth nonvanishing factor. The
asymptotic continuation used in the main text is that this smooth dressing
renormalizes only the coefficient of
Eq.~\eqref{eq:sm-canonical-linear}, not its power:
\begin{equation}
\bm\eta^{\mathsf T}T_r^{-1}\bm\eta
=
-\kappa r+o(r),
\qquad
0<|\kappa|<\infty.
\label{eq:sm-A1}
\end{equation}
Combining this with
$X_r\sim r^{-x_\sigma}$ and the exact identity
\eqref{eq:sm-K-cofactor} yields
\begin{equation}
K_r\sim \mathcal D_K\,r^{1-x_\sigma}.
\end{equation}
Equation~\eqref{eq:sm-A1} is therefore exact in the canonical singular
sector and is the first asymptotic continuation used for the fully
optimized long-range symbol.

\subsection{Momentum-resolved cofactor scaling}

Define the momentum-resolved cofactor sum
\begin{equation}
\mathcal K_r(k)
=
(-1)^r
\sum_{a,b=1}^r
\Delta_{ab}^{(r)}
(-1)^{b-a+1}
e^{-ik(b-a+1)},
\qquad
\mathcal K_r(0)=K_r .
\label{eq:sm-Kcal-def}
\end{equation}
For the canonical symbol \eqref{eq:sm-canonical-symbol}, the Cauchy
representation gives an exact expression for the normalized kernel:
\begin{equation}
\frac{\mathcal K_r(k)}{K_r}
=
e^{-ik}
\frac{
\bm v_-^{\mathsf T}(k)\,
C_\lambda^{-1}\,
\bm v_+(k)
}{
r\lambda
},
\qquad
[\bm v_\pm(k)]_a=e^{\pm ika}.
\label{eq:sm-Kcal-Cauchy}
\end{equation}
Thus the joint large-distance/small-momentum limit is reduced to the
finite-section asymptotics of an explicit Cauchy inverse.

The $kr$ structure can already be seen directly from
Eq.~\eqref{eq:sm-Kcal-Cauchy}. Setting $k=u/r$, the oscillatory factors
depend on the scaled coordinates $a/r$ and $b/r$ only. Inserting the exact
product form of $C_\lambda^{-1}$ and applying Stirling asymptotics to the
resulting products converts the normalized bilinear form into a finite
continuum integral. Consequently,
\begin{equation}
\lim_{\substack{r\to\infty\\ k\to0\\ kr=u}}
\frac{\mathcal K_r(k)}{K_r}
=
\Psi_\lambda(u),
\qquad
\Psi_\lambda(0)=1 .
\label{eq:sm-Psi-scaling}
\end{equation}
The smooth factor dressing the canonical jump is assumed not to change the
scaling variable $u=kr$; this is the second asymptotic continuation used in
the main text.

An exact finite-$r$ identity fixes the first derivative at the origin.
Let
\begin{equation}
N=\operatorname{diag}(1,\ldots,r).
\end{equation}
The Cauchy matrix obeys
\begin{equation}
C_\lambda N-NC_\lambda+\lambda C_\lambda
=
\bm 1\bm 1^{\mathsf T}.
\label{eq:sm-Cauchy-commutator}
\end{equation}
Multiplying by $C_\lambda^{-1}$ and using
Eq.~\eqref{eq:sm-Cauchy-sum} in the derivative of
Eq.~\eqref{eq:sm-Kcal-Cauchy} gives
\begin{equation}
\left.
\frac{d}{dk}
\frac{\mathcal K_r(k)}{K_r}
\right|_{k=0}
=
-i\,[1+\lambda(r-1)] .
\label{eq:sm-Kcal-derivative}
\end{equation}
Hence, in the scaling limit $k=u/r$,
\begin{equation}
\Psi_\lambda(u)
=
1-i\lambda u+O(u^2).
\label{eq:sm-Psi-small-u}
\end{equation}
The leading regular correction is therefore explicitly linear and has
nonzero coefficient in the canonical single-jump sector. For
$\lambda=1/2$,
Eq.~\eqref{eq:sm-Kcal-derivative} reduces to
\begin{equation}
\mathcal K_r^{(0)\prime}(0)
=
-\frac{i(r+1)}{2}\,K_r^{(0)},
\end{equation}
which is the exact short-range check quoted in the main analysis.

Using
$K_r\sim\mathcal D_K r^{1-x_\sigma}$, define
\begin{equation}
\mu=\alpha+x_\sigma-2.
\label{eq:sm-mu-def}
\end{equation}
The long-range part of the subtracted Euler field then has the asymptotic
form
\begin{equation}
\mathcal F_E^{\rm LR}(\pi-k)
\sim
J\mathcal D_K
\sum_{r\ge1}
\frac{1-\Psi(kr)}{r^{1+\mu}} .
\label{eq:sm-Mellin-sum}
\end{equation}
For $0<\mu<1$, setting $u=|k|r$ gives the Mellin form
\begin{equation}
\mathcal F_E^{\rm LR}(\pi-k)
\sim
J\mathcal D_K |k|^\mu
\int_0^\infty du\,
u^{-1-\mu}[1-\Psi_\pm(u)] .
\label{eq:sm-Mellin-integral}
\end{equation}
Here $\Psi_\pm$ denotes the appropriate side of the endpoint. The lower
bound $\mu>0$ ensures convergence of the large-$u$ tail when the scaling
kernel is bounded. From Eq.~\eqref{eq:sm-Psi-small-u},
$1-\Psi(u)=O(u)$, so near the origin the integrand behaves as
$u^{-\mu}$ and is integrable precisely for $\mu<1$. Therefore, provided
the Mellin amplitude is nonzero,
\begin{equation}
\left|
\mathcal F_E^{\rm LR}(\pi-k)
\right|
\sim
|k|^\mu,
\qquad
0<\mu<1 .
\label{eq:sm-LR-power}
\end{equation}

For $\mu>1$, the first Taylor moment is convergent and an ordinary linear
term in $k$ is allowed and dominates over $|k|^\mu$. At $\mu=1$, the
borderline harmonic sum gives the usual possibility of a
$k\log|k|$ correction. Thus, in the absence of an exact cancellation
between the regular and long-range amplitudes,
\begin{equation}
\zeta_E
=
\min\{\alpha+x_\sigma-2,1\}.
\label{eq:sm-zeta-min}
\end{equation}
The nonvanishing of the relevant amplitude and the absence of a fine-tuned
cancellation are the associated nondegeneracy conditions.

\subsection{Fisher--Hartwig jump and the residual phase closure}

The final step concerns the phase of the Euler field. Since
\begin{equation}
G_c(q)
=
-\frac{\mathcal F_E(q)}{|\mathcal F_E(q)|},
\label{eq:sm-G-from-F}
\end{equation}
the phase of the critical Gaussian symbol is fixed by the phase, rather
than the magnitude, of the Euler field. This distinction is important:
the exponent $\zeta_E$ defined from
$|\mathcal F_E(\pi-k)|$ does not by itself determine the jump of $G_c$.

The winding structure of the Toeplitz determinant can nevertheless be
made explicit. Because
$[T_r]_{ab}=g_{b-a+1}$, the Toeplitz symbol entering the determinant is
the shifted symbol
\begin{equation}
\tau(q)=e^{-iq}G_c(q).
\label{eq:sm-shifted-symbol}
\end{equation}
For the canonical family \eqref{eq:sm-canonical-symbol},
\begin{equation}
\tau_s(q)
=
-e^{-i(1-s)q}
=
-e^{-i\lambda q}.
\end{equation}
Thus the single Fisher--Hartwig jump lies in the winding branch
\begin{equation}
\delta=\lambda=1-s,
\label{eq:sm-delta-s}
\end{equation}
and its Toeplitz determinant has the power
\begin{equation}
x_\sigma=\delta^2 .
\label{eq:sm-FH-x}
\end{equation}
Equations~\eqref{eq:sm-delta-s} and \eqref{eq:sm-FH-x} are exact within
the canonical single-jump representation. In terms of the endpoint phase
$\varphi_{\pi^-}=s\pi$,
\begin{equation}
\delta
=
1-\frac{\varphi_{\pi^-}}{\pi}.
\label{eq:sm-delta-endpoint}
\end{equation}

What remains is to relate $\varphi_{\pi^-}$ to the magnitude exponent
$\zeta_E$. If the complex endpoint asymptotic of the Euler field is
phase locked to
\begin{equation}
\mathcal F_E(\pi-k)
=
C\,|k|^{\zeta_E}
e^{i\pi\zeta_E/2}
[1+o(1)],
\qquad
k\downarrow0,
\qquad C>0,
\label{eq:sm-phase-lock}
\end{equation}
with the opposite side fixed by the reality symmetry, then
Eq.~\eqref{eq:sm-G-from-F} gives
\begin{equation}
\frac{\varphi_{\pi^-}}{\pi}
=
\frac{\zeta_E}{2}.
\end{equation}
Combining this with Eq.~\eqref{eq:sm-delta-endpoint} yields
\begin{equation}
\delta=1-\frac{\zeta_E}{2},
\qquad
x_\sigma=\delta^2.
\label{eq:sm-phase-closure}
\end{equation}

At the short-range Ising point,
$\zeta_E=1$ and
$\varphi_{\pi^-}=\pi/2$, so
Eq.~\eqref{eq:sm-phase-lock} and
Eq.~\eqref{eq:sm-phase-closure} are exact and give
$\delta=1/2$ and $x_\sigma=1/4$. The single-jump winding relation
\eqref{eq:sm-delta-endpoint} is also exact throughout the canonical family.
The genuinely residual assumption is therefore the phase-locking
continuation \eqref{eq:sm-phase-lock} for the fully optimized long-range
Euler field. It is stronger than the magnitude statement
$|\mathcal F_E|\sim|k|^{\zeta_E}$ and should not be inferred from that
magnitude alone.

With this phase closure, the long-range branch satisfies
\begin{equation}
x_\sigma
=
\left[
1-\frac{\alpha+x_\sigma-2}{2}
\right]^2,
\end{equation}
whose continuously connected solution is
\begin{equation}
x_\sigma^{GA}
=
\left(\sqrt{5-\alpha}-1\right)^2 .
\end{equation}
When the long-range power reaches the regular linear contribution,
$\zeta_E=1$, the Fisher--Hartwig closure gives
$x_\sigma=1/4$ and hence
\begin{equation}
\alpha_\star=\frac{11}{4}.
\end{equation}
Thus the value $11/4$ is not inserted as a separate crossover assumption;
it follows internally from the long-range power counting together with
the single-jump phase closure. The latter is the only one of the three
ingredients for which the present argument remains a phase-locking
continuation rather than an exact result for the fully optimized
long-range symbol.

\section{Logical status of the analytic critical relations}
\label{sm:status-critical-relations}

It is useful to distinguish three logically different levels in the
critical analysis: exact statements within the Gaussian variational
manifold, asymptotic closures used to obtain closed analytic formulas, and
statements about the infrared fixed point of the microscopic interacting
spin chain.

\paragraph{Exact Gaussian identities versus asymptotic closures.}
The variational energy functional, the Pfaffian/Toeplitz expressions for
spin correlations, the cofactor identities, and the stationarity equation
\begin{equation}
G(q)=-\frac{\mathcal F_E(q)}{|\mathcal F_E(q)|}
\label{eq:sm-status-stationarity}
\end{equation}
are exact within the chosen Gaussian manifold. Likewise, on the continuous
ferromagnetic branch connected to the short-range solution,
$\mathcal F_E(\pi)=0$ is the exact Gaussian criticality condition. These
statements do not rely on replacing the interacting spin Hamiltonian by a
quadratic Hamiltonian. At the nearest-neighbor endpoint, the Gaussian
manifold contains the exact free-fermion ground state \cite{Pfeuty1970}.

The closed critical exponents require additional asymptotic information.
As discussed in Sec.~\ref{sm:analytic-closures}, the derivation of
\begin{equation}
\zeta_E=\min\{\alpha+x_\sigma-2,1\}
\label{eq:sm-status-zeta}
\end{equation}
uses the large-$r$ inverse-Toeplitz behavior and the scaling form of the
momentum-resolved cofactor kernel. The Fisher--Hartwig result
$x_\sigma=\delta^2$ is exact once the critical symbol belongs to the
corresponding single-jump class
\cite{JinKorepin2004,ItsJinKorepin2005,AresEtAl2015,AresEtAl2018},
whereas the relation
\begin{equation}
\delta=1-\frac{\zeta_E}{2}
\label{eq:sm-status-phase}
\end{equation}
requires the additional phase-locking continuation discussed there.
Accordingly, the closed expression for $x_\sigma^{GA}$ should be understood
as the analytic solution of this Gaussian asymptotic closure.

\paragraph{Long-range Gaussian fixed point.}
Writing $\sigma=\alpha-1$, the conventional continuum description of the
long-range Ising transition contains the quadratic kernel
\begin{equation}
S_2\sim
\frac12\int d\omega\,dk\,
\bigl[\omega^2+c|k|^\sigma+r\bigr]
|\phi(k,\omega)|^2 .
\label{eq:sm-status-action}
\end{equation}
Such long-range quantum fixed-point descriptions have been developed in
Refs.~\cite{DuttaBhattacharjee2001,Maghrebi2016,Defenu2017,DefenuRMP2023}.
If the microscopic chain flows to this Gaussian long-range fixed point,
balancing frequency and momentum gives
\begin{equation}
z=\frac{\sigma}{2},
\qquad
x_\sigma=1-\frac{\sigma}{2}
\qquad (d=1).
\label{eq:sm-status-mf}
\end{equation}
Power counting gives the quartic coupling the scaling dimension
$y_u=3\sigma/2-d$, so that it is irrelevant for
$\sigma<2/3$ in one dimension
\cite{FisherMaNickel1972,DuttaBhattacharjee2001,Maghrebi2016,Defenu2017,DefenuRMP2023}.
Related rigorous mean-field and Gaussian results are known for classical
long-range Ising models and for high-dimensional quantum Ising systems
\cite{AizenmanFernandez1988,ChenSakai2015,Bjornberg2013,KamijimaSakai2025},
although these results do not directly prove the infrared behavior of the
present one-dimensional power-law quantum chain. These exponents are
therefore properties of the Gaussian long-range fixed point, while their
application to the microscopic long-range quantum Ising chain assumes that
the lattice model flows to this infrared theory. Dangerous irrelevant
variables can also complicate ordinary finite-size scaling in this regime
\cite{LangheldKoziolAdelhardtKapferSchmidt2022,Koziol2021,ShirataniTodo2024}.

The Gaussian variational result is nevertheless very close to the
conventional fixed-point form. Expanding the closed expression at small
$\sigma$ gives
\begin{equation}
x_\sigma^{GA}
=
1-\frac{\sigma}{2}
+\frac{\sigma^2}{32}
+O(\sigma^3),
\label{eq:sm-status-expansion}
\end{equation}
so the two agree through first order and differ only from
$O(\sigma^2)$ onward.

\paragraph{Euler exponent and physical dynamical exponent.}
The physical long-range scaling relation
\begin{equation}
z=\alpha+x_\sigma-2
\label{eq:sm-status-zphys}
\end{equation}
is logically distinct from the Gaussian Euler relation
\begin{equation}
\zeta_E^{GA}
=
\alpha+x_\sigma^{GA}-2
\label{eq:sm-status-zetaGA}
\end{equation}
on the nonanalytic branch. The former follows from the long-range fixed
point and its nonanalytic spatial kernel
\cite{Maghrebi2016,Defenu2017,DefenuRMP2023}, whereas the latter follows
from the Gaussian stationarity equation. Consequently, literal equality
of the physical $z$ and the Gaussian $\zeta_E$ requires, in addition, that
the corresponding magnetic exponents agree. The comparison in
Fig.~\ref{fig:hc_exponent_main}(b) is therefore best interpreted as a
numerical test of how accurately the static Gaussian Euler singularity
tracks the physical dynamical exponent, rather than as an identification
of $\mathcal F_E$ with an excitation spectrum. Existing QMC results provide
a direct benchmark for this comparison
\cite{Koziol2021,ShirataniTodo2024}.

\paragraph{Long-range--to--short-range crossover.}
Within the Gaussian asymptotic closure, the crossover occurs when the
nonanalytic Euler contribution reaches the regular linear term,
$\alpha+x_\sigma-2=1$. Combining this with the single-jump phase closure
gives
\begin{equation}
\alpha_\star=\frac{11}{4}.
\label{eq:sm-status-crossover}
\end{equation}
Writing $\sigma_\star=\alpha_\star-1$, this is
$\sigma_\star=7/4=2-\eta_{\rm SR}$ with
$\eta_{\rm SR}=1/4$, i.e. the Sak criterion \cite{Sak1973}. This shifted
boundary differs from the unshifted $\sigma_\star=2$ continuation of the
original long-range RG picture \cite{FisherMaNickel1972}. Thus the Gaussian
asymptotic theory naturally reproduces the Sak boundary; the value $11/4$
is not inserted as a separate assumption. Its physical interpretation
should nevertheless be separated from the internal Gaussian derivation,
since the precise long-range--to--short-range crossover remains unsettled.
Numerical and field-theory studies have supported the Sak scenario
\cite{HoritaSuwaTodo2017,BehanRastelliRychkovZan2017,ShirataniTodo2024},
while recent work has also explored alternative crossover structures
\cite{XiaoIsing2025,LiQuantumRG2026}.

\paragraph{Entanglement coefficient.}
A similar distinction applies to the logarithmic entanglement coefficient.
The entropy obtained directly from the restricted covariance matrix is
exact for the optimized Gaussian state
\cite{Peschel2003,PeschelEisler2009}. The closed thermodynamic formula for
$c_{\rm eff}$ additionally requires the spectral-class identification
described in the entanglement-spectrum closure. The relation between
singular fermionic symbols, Toeplitz asymptotics, and logarithmic
entanglement is well established
\cite{JinKorepin2004,ItsJinKorepin2005,AresEtAl2015,AresEtAl2018}.
Once the additional spectral identification is made, the remaining
free-fermion spectral and dilogarithm formulas are the known exact results
of Refs.~\cite{EislerPeschel2010,PeschelEisler2012}; the nontrivial step in
the present theory is the identification of the optimized long-range
covariance spectrum with that reference spectral class.

The overall conclusion is therefore independent of claiming exactness of
every closed analytic expression: the Gaussian variational manifold
captures the singular organization of the critical theory much more
accurately than the global wave-function fidelity alone would suggest.

\section{Complete Gaussian/Pfaffian variational manifold}\label{sm:full-gaussian}
\label{sec:ansatz}

We first formulate the variational construction for a general pure Gaussian
state and an arbitrary reference Fock configuration.  The real,
translation-invariant manifold used for the thermodynamic calculations in
the main text is obtained as a specialization of this general framework.

\subsection{Gaussian pairing state and normalization}
\label{subsec:gaussian_state}

We consider a fermionic Gaussian state built on an arbitrary fermionic occupation
configuration of $L$ spinless modes. Let $c_i^\dagger$ and $c_i$
($i=1,\dots,L$) denote the canonical fermionic creation and annihilation
operators, satisfying
\begin{equation}
\{c_i,c_j^\dagger\}=\delta_{ij},
\qquad
\{c_i,c_j\}=\{c_i^\dagger,c_j^\dagger\}=0.
\end{equation}
We choose a reference Fock configuration
\begin{equation}
\ket{\mathcal C}
=
\ket{n_1,\dots,n_L},
\qquad
n_i\in\{0,1\}.
\label{eq:reference-configuration}
\end{equation}
and introduce the particle--hole transformed fermionic operators
\begin{equation}
d_i=(1-n_i)c_i+n_i c_i^\dagger,
\qquad
d_i^\dagger=(1-n_i)c_i^\dagger+n_i c_i.
\end{equation}
By construction,
\begin{equation}
d_i\ket{\mathcal C}=0,
\qquad
i=1,\dots,L,
\end{equation}
so that $\ket{\mathcal C}$ plays the role of the fermionic vacuum for the
$d_i$ modes.

Starting from the reference configuration $\ket{\mathcal C}$, we introduce
the variational Gaussian pairing state
\begin{equation}
\label{eq:psiR}
\ket{\RR,\mathcal C}
=
\frac{1}{\mathcal{N}_{R}}
\exp\!\left(
\frac{1}{2}\sum_{i,j=1}^{L}
R_{ij}\,d_i^\dagger d_j^\dagger
\right)
\ket{\mathcal C}.
\end{equation}
Here $\RR$ is an $L\times L$ antisymmetric matrix, which is the natural structure required for fermionic pairing. Equivalently, this state admits the computational-basis expansion
\begin{equation}
\label{eq:psiR_pfaffian}
\ket{\RR,\mathcal C}
=
\frac{1}{\mathcal N_R}
\sum_{\substack{I}}
\operatorname{sgn}(\mathcal C,I)\,
\operatorname{pf}\!\left(\RR_{I(\mathcal C)}\right)
\ket{I},
\end{equation}
where
\begin{equation}
\ket{I}=\ket{m_1,\dots,m_L},
\qquad
I(\mathcal C)=\{i\mid n_i\neq m_i\},
\end{equation}
and $\RR_{I(\mathcal C)}$ denotes the principal submatrix of $\RR$ restricted
to the sites in $I(\mathcal C)$. The fermionic reordering sign is
\begin{equation}
\label{eq:configuration_sign}
\operatorname{sgn}(\mathcal C,I)
=
\prod_{i=2}^{L}
(-1)^{
|n_i-m_i|
\sum_{j=1}^{i-1}n_j
}.
\end{equation}
We use the convention $\operatorname{pf}(\varnothing)=1$. Since the exponential changes the occupation relative to $\ket{\mathcal C}$
only in pairs, the state $\ket{\RR,\mathcal C}$ has definite fermion parity,
equal to that of the reference configuration:
\begin{equation}
\hat{\mathcal P}\ket{\RR,\mathcal C}
=
(-1)^{N_{\mathcal C}}
\ket{\RR,\mathcal C},
\qquad
N_{\mathcal C}=\sum_{i=1}^{L}n_i.
\end{equation}
In particular, the state has even parity when $N_{\mathcal C}$ is even and
odd parity when $N_{\mathcal C}$ is odd.

The normalization factor is
\begin{equation}
\label{eq:norm}
\mathcal{N}_{R}
=
\det(\Id+\RR^{\dagger}\RR)^{1/4},
\end{equation}
For real antisymmetric $\RR$, one has
\begin{equation}
\det(\Id+\RR^\dagger \RR)
=
\det(\Id-\RR^2)
=
\det(\Id-\RR)^2.
\end{equation}
This state provides the variational manifold used in the following and is
characterized by the antisymmetric pairing matrix $\RR$ together with the
reference configuration $\mathcal C$. The vacuum-reference state
$\ket{\RR,0}$ is recovered by choosing $n_i=0$ for all $i$, in which case
$\operatorname{sgn}(0,I)=1$.

\subsection{Basic fermionic correlators}
\label{subsec:ferm_corrs}

The Gaussian pairing state~\eqref{eq:psiR} is completely characterized by
its equal-time two-point functions. We introduce the normal and anomalous
correlation matrices of the particle--hole transformed fermions,
\begin{align}
\label{eq:C_def}
C_{jk}(\RR,\mathcal C)
&=
\bra{\RR,\mathcal C}
d_j^\dagger d_k
\ket{\RR,\mathcal C},
\\
\label{eq:F_def}
F_{jk}(\RR,\mathcal C)
&=
\bra{\RR,\mathcal C}
d_j^\dagger d_k^\dagger
\ket{\RR,\mathcal C},
\end{align}
which we collect into $\mathbf C$ and $\mathbf F$. For the
state~\eqref{eq:psiR}, one may express these correlators directly in terms
of the pairing matrix $\RR$ by defining
\begin{equation}
\mathbf Q=(\Id+\RR^\dagger\RR)^{-1}.
\end{equation}
A standard calculation then yields
\begin{equation}
\label{eq:CF_in_terms_of_R}
\mathbf C=\Id-\mathbf Q,
\qquad
\mathbf F=\RR^*\mathbf Q^T,
\end{equation}
so that $\mathbf C$ is Hermitian,
while $\mathbf F$ is antisymmetric, as required by fermionic statistics.
To connect with Jordan--Wigner strings, it is convenient to work with the
Majorana-like combinations
\begin{equation}
\gamma_j=d_j+d_j^\dagger,
\qquad
\bar{\gamma}_j=d_j^\dagger-d_j,
\end{equation}
and the corresponding two-point functions
\begin{align}
\label{eq:GK_defs}
G_{jk}(\RR,\mathcal C)
&=
\bra{\RR,\mathcal C}
\bar{\gamma}_j\gamma_k
\ket{\RR,\mathcal C},
\\
K_{jk}(\RR,\mathcal C)
&=
\bra{\RR,\mathcal C}
\gamma_j\gamma_k
\ket{\RR,\mathcal C},
\\
\bar{K}_{jk}(\RR,\mathcal C)
&=
\bra{\RR,\mathcal C}
\bar{\gamma}_j\bar{\gamma}_k
\ket{\RR,\mathcal C}.
\end{align}
In terms of $\mathbf C$ and $\mathbf F$, these matrices can be written
compactly as
\begin{align}
\label{eq:GK_in_terms_of_CF}
\mathbf G
&=
\mathbf F-\mathbf F^\dagger
+\mathbf C+\mathbf C^T-\Id,
\\
\mathbf K
&=
\mathbf F+\mathbf F^\dagger
+\mathbf C-\mathbf C^T+\Id,
\\
\bar{\mathbf K}
&=
\mathbf F+\mathbf F^\dagger
-\mathbf C+\mathbf C^T-\Id.
\end{align}
These objects appear naturally in Pfaffian and determinant expressions for
spin correlators in Sec.~\ref{sec:spin_corrs}. In the common case where the
variational optimum can be chosen with real $\RR$, one has
\begin{equation}
\mathbf K=\Id,
\qquad
\bar{\mathbf K}=-\Id.
\end{equation}
Therefore, the off-diagonal elements of both $\mathbf K$ and
$\bar{\mathbf K}$ vanish, and the remaining nontrivial information relevant
for $\sigma^x$ strings is encoded in $\mathbf G$.

The same contractions also determine the Majorana covariance matrix used
below for subsystem entanglement. Since $\gamma_j$ is Hermitian whereas
$\bar{\gamma}_j$ is anti-Hermitian, we introduce the Hermitian Majorana
vector
$\mathbf a=(\gamma_1,\ldots,\gamma_L,
i\bar{\gamma}_1,\ldots,i\bar{\gamma}_L)^T$, whose components satisfy
$\{a_\mu,a_\nu\}=2\delta_{\mu\nu}$.  The covariance matrix is defined by
\begin{equation}
\label{eq:covariance_def}
\Gamma_{\mu\nu}
=
\frac{i}{2}
\bra{\RR,\mathcal C}
[a_\mu,a_\nu]
\ket{\RR,\mathcal C},
\end{equation}
or equivalently
$\langle a_\mu a_\nu\rangle=\delta_{\mu\nu}-i\Gamma_{\mu\nu}$.
In terms of the contractions introduced above,
\begin{equation}
\label{eq:covariance_GK}
\boldsymbol{\Gamma}
=
\begin{pmatrix}
i(\mathbf K-\Id) & \mathbf G^T\\
-\mathbf G & -i(\bar{\mathbf K}+\Id)
\end{pmatrix}.
\end{equation}
For real antisymmetric $\RR$, where
$\mathbf K=\Id$ and $\bar{\mathbf K}=-\Id$, this reduces to
$\boldsymbol{\Gamma}
=\bigl(\begin{smallmatrix}
0&\mathbf G^T\\[-1pt]-\mathbf G&0
\end{smallmatrix}\bigr)$.
For a pure Gaussian state,
$\boldsymbol{\Gamma}^T=-\boldsymbol{\Gamma}$ and
$\boldsymbol{\Gamma}^2=-\Id_{2L}$.

\section{Jordan--Wigner representation and spin correlators}
\label{sm:jw-route}
\label{sec:spin_corrs}

We derive here the spin correlators required in the Gaussian
calculations. After fixing the Jordan--Wigner and particle--hole
conventions, we express the observables in terms of the Gaussian two-point
matrices $\mathbf C$, $\mathbf F$, $\mathbf G$, $\mathbf K$, and
$\bar{\mathbf K}$. For real antisymmetric $\mathbf R$, the relevant
Pfaffian expressions reduce to determinants.

We map spins to the physical fermions according to
\begin{equation}
\label{eq:JW}
c_l
=
\left(
\prod_{m<l}(-\sigma_m^z)
\right)\sigma_l^-,
\qquad
c_l^\dagger
=
\left(
\prod_{m<l}(-\sigma_m^z)
\right)\sigma_l^+,
\end{equation}
so that
\begin{equation}
\sigma_l^z=2c_l^\dagger c_l-1,
\qquad
\sigma_l^x
=
\left(
\prod_{m<l}(-\sigma_m^z)
\right)
(c_l+c_l^\dagger).
\end{equation}
The physical fermions and the particle--hole transformed fermions are
related by
\begin{equation}
c_l=(1-n_l)d_l+n_l d_l^\dagger,
\qquad
c_l^\dagger=(1-n_l)d_l^\dagger+n_l d_l .
\end{equation}
Introducing $s_l=1-2n_l$,
one obtains
\begin{equation}
c_l+c_l^\dagger=\gamma_l,
\qquad
c_l^\dagger-c_l=s_l\bar{\gamma}_l,
\end{equation}
and therefore
\begin{equation}
\label{eq:sz_d}
\sigma_l^z
=
s_l\left(2d_l^\dagger d_l-1\right).
\end{equation}
To keep track of the signs associated with the reference configuration, we introduce the diagonal matrix
\begin{equation}
\mathbf S_{\mathcal C}
=
\operatorname{diag}(s_1,\dots,s_L),
\qquad
\mathbf S_{\mathcal C}^2=\Id.
\end{equation}
For later use, for $l<m$ we also define the reference-configuration string
factor
\begin{equation}
\chi_{lm}
=
\prod_{p=l}^{m-1}s_p
=
(-1)^{\sum_{p=l}^{m-1}n_p}.
\label{eq:chi-lm}
\end{equation}

\subsection{Correlators from fermionic two-point matrices}
\label{subsec:spin_fermionic}
\label{sm:ferm-route}

Here, all spin observables are evaluated from the Gaussian
two-point functions.  Longitudinal correlators follow directly from
$\mathbf C$ and $\mathbf F$, whereas transverse Jordan--Wigner strings are
expressed through the Majorana-bilinear contractions
$\mathbf G$, $\mathbf K$, and $\bar{\mathbf K}$ defined in
Sec.~\ref{subsec:ferm_corrs}.  For real $\mathbf R$,
$\mathbf K=\mathbf I$ and $\bar{\mathbf K}=-\mathbf I$, and the relevant
Pfaffians reduce to determinants involving only $\mathbf G$.
A compact summary of the resulting one-, two-, three-, and four-body
correlator formulas is given in Table~\ref{tab:spin_ferm_route_realR}.

\subsubsection{One-body longitudinal correlator}
\label{subsubsec:sz_ferm}

Because $\ket{\RR,\mathcal C}$ has definite fermion parity,
\begin{equation}
\label{eq:sx_vanish}
\langle\sigma_l^x\rangle_{\RR,\mathcal C}=0.
\end{equation}

Using Eq.~\eqref{eq:sz_d},
\begin{equation}
\label{eq:sz_C}
\langle\sigma_l^z\rangle_{\RR,\mathcal C}
=
s_l\left(2C_{ll}-1\right).
\end{equation}

\subsubsection{Two-body longitudinal correlator}
\label{subsubsec:szz_ferm}

For $\langle\sigma_l^z\sigma_m^z\rangle_{\RR,\mathcal C}$, Wick's theorem
gives
\begin{equation}
\label{eq:szz_CF}
\begin{split}
\langle\sigma_l^z\sigma_m^z\rangle_{\RR,\mathcal C}
={}&\,
s_ls_m
\Big[
4C_{ll}C_{mm}
+4C_{lm}\delta_{lm}
-4C_{lm}C_{ml}
\\
&\hspace{13mm}
-4F_{lm}F^*_{ml}
-2C_{ll}
-2C_{mm}
+1
\Big].
\end{split}
\end{equation}

To write the sum over distinct pairs compactly, define
\begin{equation}
t_{\mathcal C}
=
\operatorname{Tr}\mathbf S_{\mathcal C}
=
\sum_{j=1}^{L}s_j .
\end{equation}
Summing Eq.~\eqref{eq:szz_CF} over $l<m$ yields
\begin{equation}
\label{eq:szz_sum}
\begin{split}
\sum_{l<m}
\langle\sigma_l^z\sigma_m^z\rangle_{\RR,\mathcal C}
={}&\,
2\left[
\operatorname{Tr}
\left(
\mathbf S_{\mathcal C}\mathbf C
\right)
\right]^2
-2\operatorname{Tr}
\left(
\mathbf S_{\mathcal C}\mathbf C
\mathbf S_{\mathcal C}\mathbf C
\right)
\\
&-
2\operatorname{Tr}
\left(
\mathbf S_{\mathcal C}\mathbf F
\mathbf S_{\mathcal C}\mathbf F^*
\right)
-2t_{\mathcal C}
\operatorname{Tr}
\left(
\mathbf S_{\mathcal C}\mathbf C
\right)
\\
&+
2\operatorname{Tr}\mathbf C
+\frac{t_{\mathcal C}^{\,2}-L}{2}.
\end{split}
\end{equation}

For the vacuum reference,
$\mathbf S_{\mathcal C}=\mathbf I$ and $t_{\mathcal C}=L$, giving
\begin{equation}
\begin{split}
\sum_{l<m}
\langle\sigma_l^z\sigma_m^z\rangle_{\RR,0}
={}&\,
2\big(\operatorname{Tr}\mathbf C\big)^2
-2\operatorname{Tr}\big(\mathbf C^2\big)
+2(1-L)\operatorname{Tr}\mathbf C
\\
&-
2\operatorname{Tr}\big(\mathbf F\mathbf F^*\big)
+\frac{L(L-1)}{2}.
\end{split}
\end{equation}

\begin{table*}[t]
\centering
\small
\renewcommand{\arraystretch}{1.42}
\setlength{\tabcolsep}{5pt}
\rowcolors{2}{PDAstripe}{white}
\begin{tabular}{lll}
\rowcolor{PDAblue}
\parbox[c]{0.235\textwidth}{\centering\color{white}\bfseries Correlator} &
\parbox[c]{0.135\textwidth}{\centering\color{white}\bfseries Conditions} &
\parbox[c]{0.57\textwidth}{\centering\color{white}\bfseries Real-$\mathbf R$ expression} \\
\midrule
\parbox[t]{0.235\textwidth}{$\langle\sigma_j^x\rangle$} &
\parbox[t]{0.135\textwidth}{\centering definite parity} &
\parbox[t]{0.57\textwidth}{$0$} \\
\parbox[t]{0.235\textwidth}{$\langle\sigma_j^z\rangle$} &
\parbox[t]{0.135\textwidth}{\centering any $j$} &
\parbox[t]{0.57\textwidth}{$s_j(2C_{jj}-1)$} \\
\parbox[t]{0.235\textwidth}{$\langle\sigma_l^z\sigma_m^z\rangle$} &
\parbox[t]{0.135\textwidth}{\centering any $l,m$} &
\parbox[t]{0.57\textwidth}{$\displaystyle s_ls_m\!\left[4C_{ll}C_{mm}+4C_{lm}\delta_{lm}-4C_{lm}C_{ml}-4F_{lm}F^*_{ml}-2C_{ll}-2C_{mm}+1\right]$} \\
\parbox[t]{0.235\textwidth}{$\langle\sigma_l^x\sigma_m^x\rangle$} &
\parbox[t]{0.135\textwidth}{\centering $l<m$} &
\parbox[t]{0.57\textwidth}{$\displaystyle \chi_{lm}\det\widetilde{\boldsymbol\Gamma}^{xx}(l,m)$} \\
\parbox[t]{0.235\textwidth}{$\langle\sigma_l^x\sigma_m^x\sigma_n^z\rangle$} &
\parbox[t]{0.135\textwidth}{\centering $n<l$ or $n>m$} &
\parbox[t]{0.57\textwidth}{$\displaystyle \chi_{lm}s_n\det\widetilde{\boldsymbol\Gamma}^{xxz}(l,m;n)$} \\
\parbox[t]{0.235\textwidth}{$\langle\sigma_l^x\sigma_m^x\sigma_n^z\rangle$} &
\parbox[t]{0.135\textwidth}{\centering $l<n<m$} &
\parbox[t]{0.57\textwidth}{$\displaystyle -\chi_{lm}s_n\det\!\left[\widetilde{\boldsymbol\Gamma}^{xx}_{\overline{n-l+1},\,\overline{n-l}}(l,m)\right]$} \\
\parbox[t]{0.235\textwidth}{$\langle\sigma_l^x\sigma_m^x\sigma_n^x\sigma_j^x\rangle$} &
\parbox[t]{0.135\textwidth}{\centering $l<m<n<j$} &
\parbox[t]{0.57\textwidth}{$\displaystyle \chi_{lm}\chi_{nj}\det\widetilde{\boldsymbol\Gamma}^{xxxx}(l,m;n,j)$} \\
\bottomrule
\end{tabular}
\caption{Checked determinant formulas used for real $\mathbf R$. The inside-interval three-point formula uses one-based matrix notation for the deleted row and column.}
\label{tab:spin_ferm_route_realR}
\end{table*}

\subsubsection{Two-body transverse correlator}
\label{subsubsec:sxx_ferm}

The Jordan--Wigner string in $\sigma_l^x\sigma_m^x$ reduces to the
Pfaffian of a $2(m-l)\times2(m-l)$ antisymmetric matrix.  Define
$\boldsymbol{\Gamma}^{xx}(l,m)$ through its upper-triangular entries,
for $k<k'$,
\begin{equation}
\label{eq:Gamma_xx_entries}
\Gamma^{xx}_{k,k'}(l,m)
=
\begin{cases}
G_{l+\frac{k-1}{2},\,l+\frac{k'}{2}},
& k~\mathrm{odd},~k'~\mathrm{even},
\\[2pt]
-\,G_{l+\frac{k'-1}{2},\,l+\frac{k}{2}},
& k~\mathrm{even},~k'~\mathrm{odd},
\\[2pt]
\bar K_{l+\frac{k-1}{2},\,l+\frac{k'-1}{2}},
& k,k'~\mathrm{odd},
\\[2pt]
K_{l+\frac{k}{2},\,l+\frac{k'}{2}},
& k,k'~\mathrm{even}.
\end{cases}
\end{equation}
The remaining entries follow from antisymmetry.

The reference-configuration signs can be collected into
\begin{equation}
\label{eq:D_C_xx}
\mathbf D_{\mathcal C}(l,m)
=
\operatorname{diag}
\left(
s_l,1,s_{l+1},1,\ldots,s_{m-1},1
\right).
\end{equation}
Using
$\operatorname{pf}(\mathbf D\mathbf A\mathbf D^T)
=\det(\mathbf D)\operatorname{pf}(\mathbf A)$,
one finds
\begin{equation}
\label{eq:sxx_pf_Gamma}
\langle\sigma_l^x\sigma_m^x\rangle_{\RR,\mathcal C}
=
\chi_{lm}\,
\operatorname{pf}
\left(
\boldsymbol{\Gamma}^{xx}(l,m)
\right).
\end{equation}

For real $\mathbf R$, $\mathbf K=\mathbf I$ and
$\bar{\mathbf K}=-\mathbf I$, and the Pfaffian reduces to
\begin{equation}
\label{eq:sxx_det_tildeGamma}
\langle\sigma_l^x\sigma_m^x\rangle_{\RR,\mathcal C}
=
\chi_{lm}\,
\det
\left(
\widetilde{\boldsymbol{\Gamma}}^{xx}(l,m)
\right),
\end{equation}
where
\begin{equation}
\widetilde{\Gamma}^{xx}_{k,k'}(l,m)
=
G_{l+k-1,\,l+k'},
\qquad
k,k'=1,\ldots,m-l .
\end{equation}

\subsubsection{Three-body mixed correlator}
\label{subsubsec:xxz_ferm}

For distinct $l,m,n$ with $l<m$, introduce
\begin{equation}
\boldsymbol{\Gamma}^{xxz}
=
\begin{pmatrix}
\boldsymbol{\Gamma}^{xx}(l,m) & \mathbf O(l,n)\\
-\mathbf O^T(l,n) & \mathbf Z(n)
\end{pmatrix},
\end{equation}
where
\begin{equation}
\mathbf Z(n)
=
\begin{pmatrix}
0&G_{nn}\\
-G_{nn}&0
\end{pmatrix},
\end{equation}
and
\begin{equation}
O_{k,k'}(l,n)
=
\begin{cases}
G_{l+\frac{k-1}{2},\,n},
& k~\mathrm{odd},~k'~\mathrm{even},
\\[2pt]
-\,G_{n,\,l+\frac{k}{2}},
& k~\mathrm{even},~k'~\mathrm{odd},
\\[2pt]
\bar K_{l+\frac{k-1}{2},\,n},
& k,k'~\mathrm{odd},
\\[2pt]
K_{l+\frac{k}{2},\,n},
& k,k'~\mathrm{even}.
\end{cases}
\end{equation}
The general Gaussian expression is
\begin{equation}
\label{eq:xxz_RC_corr_pf}
\langle
\sigma_l^x\sigma_m^x\sigma_n^z
\rangle_{\mathbf R,\mathcal C}
=
\chi_{lm}s_n\,
\operatorname{pf}
\left(
\boldsymbol{\Gamma}^{xxz}
\right).
\end{equation}

For real $\mathbf R$, the determinant representation depends on the
location of the $z$ insertion.  If $n<l$ or $n>m$, define
\begin{equation}
\widetilde O_k(l,n)=G_{l+k-1,n},
\qquad
\widetilde W_{k'}(l,n)=G_{n,l+k'},
\end{equation}
and
\begin{equation}
\label{eq:tildeGxxz_outside}
\widetilde{\boldsymbol{\Gamma}}^{xxz}
=
\begin{pmatrix}
\widetilde{\boldsymbol{\Gamma}}^{xx}(l,m)
&
\widetilde{\mathbf O}(l,n)
\\
\widetilde{\mathbf W}^{T}(l,n)
&
G_{nn}
\end{pmatrix}.
\end{equation}
Then
\begin{equation}
\label{eq:xxz_RC_det_outside}
\langle
\sigma_l^x\sigma_m^x\sigma_n^z
\rangle_{\mathbf R,\mathcal C}
=
\chi_{lm}s_n
\det
\left(
\widetilde{\boldsymbol{\Gamma}}^{xxz}
\right).
\end{equation}

If $l<n<m$, the insertion lies inside the Jordan--Wigner interval and
\begin{equation}
\label{eq:xxz_RC_det_inside}
\langle
\sigma_l^x\sigma_m^x\sigma_n^z
\rangle_{\mathbf R,\mathcal C}
=
-\chi_{lm}s_n
\det
\left(
\widetilde{\boldsymbol{\Gamma}}^{xx}_{
\overline{\,n-l+1\,},\,\overline{\,n-l\,}
}(l,m)
\right),
\end{equation}
where the indicated row and column are removed from
$\widetilde{\boldsymbol{\Gamma}}^{xx}(l,m)$.

\subsubsection{Four-body transverse correlator}
\label{subsubsec:xxxx_ferm}

For $l<m<n<j$, define
\begin{equation}
\boldsymbol{\Gamma}^{xxxx}
=
\begin{pmatrix}
\boldsymbol{\Gamma}^{xx}(l,m) & \mathbf V(l,n)\\
-\mathbf V^T(l,n) & \boldsymbol{\Gamma}^{xx}(n,j)
\end{pmatrix},
\end{equation}
where
\begin{equation}
V_{k,k'}(l,n)
=
\begin{cases}
G_{l+\frac{k-1}{2},\,n+\frac{k'}{2}},
& k~\mathrm{odd},~k'~\mathrm{even},
\\[2pt]
-\,G_{n+\frac{k'-1}{2},\,l+\frac{k}{2}},
& k~\mathrm{even},~k'~\mathrm{odd},
\\[2pt]
\bar K_{l+\frac{k-1}{2},\,n+\frac{k'-1}{2}},
& k,k'~\mathrm{odd},
\\[2pt]
K_{l+\frac{k}{2},\,n+\frac{k'}{2}},
& k,k'~\mathrm{even}.
\end{cases}
\end{equation}
The correlator is
\begin{equation}
\label{eq:xxxx_RC_corr_pf}
\langle
\sigma_l^x\sigma_m^x\sigma_n^x\sigma_j^x
\rangle_{\mathbf R,\mathcal C}
=
\chi_{lm}\chi_{nj}\,
\operatorname{pf}
\left(
\boldsymbol{\Gamma}^{xxxx}
\right).
\end{equation}

For real $\mathbf R$, define the two cross-correlation blocks
\begin{align}
\widetilde V_{k,k'}(l,n)
&=
G_{l+k-1,\,n+k'},
&
k&=1,\ldots,m-l,
&
k'&=1,\ldots,j-n,
\\
\widetilde W_{k',k}(l,n)
&=
G_{n+k'-1,\,l+k},
&
k&=1,\ldots,m-l,
&
k'&=1,\ldots,j-n.
\end{align}
Then
\begin{equation}
\label{eq:tildeGxxxx_correct}
\widetilde{\boldsymbol{\Gamma}}^{xxxx}
=
\begin{pmatrix}
\widetilde{\boldsymbol{\Gamma}}^{xx}(l,m)
&
\widetilde{\mathbf V}(l,n)
\\
\widetilde{\mathbf W}(l,n)
&
\widetilde{\boldsymbol{\Gamma}}^{xx}(n,j)
\end{pmatrix},
\end{equation}
and
\begin{equation}
\label{eq:xxxx_RC_corr_det}
\langle
\sigma_l^x\sigma_m^x\sigma_n^x\sigma_j^x
\rangle_{\mathbf R,\mathcal C}
=
\chi_{lm}\chi_{nj}\,
\det
\left(
\widetilde{\boldsymbol{\Gamma}}^{xxxx}
\right).
\end{equation}

\begin{figure}[t]
\centering

\safeincludegraphics[width=0.72\linewidth]
{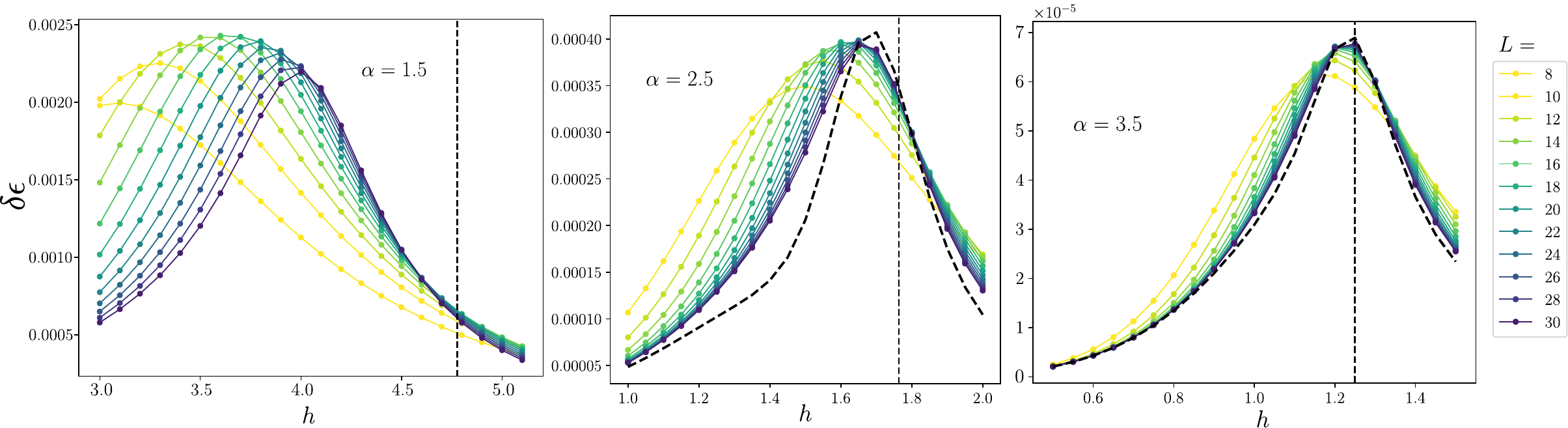}

\vspace{1mm}

\safeincludegraphics[width=0.72\linewidth]
{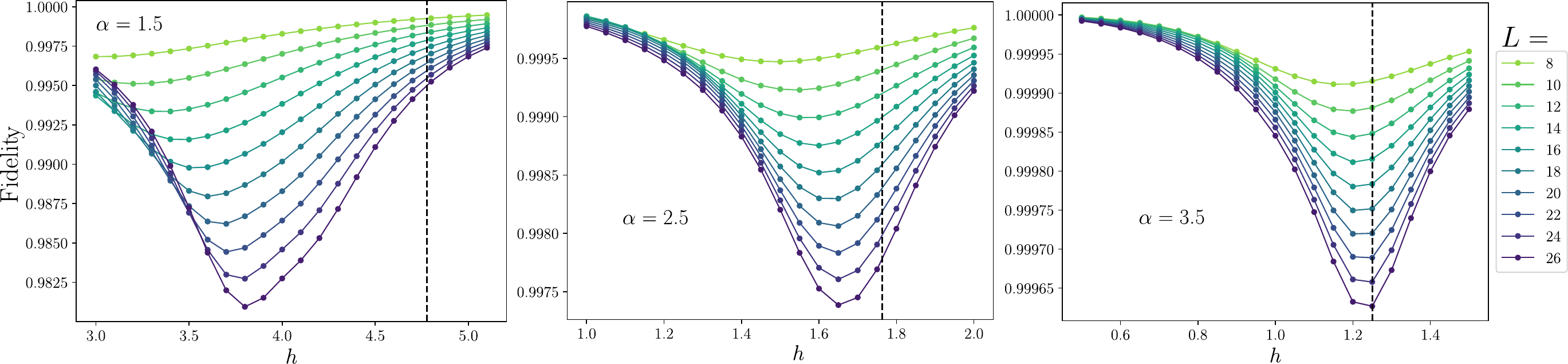}

\caption{\textbf{Finite-size variational accuracy.}
Relative energy error and fidelity of the optimized Gaussian state
compared with exact diagonalization.  The energy-error extrapolation uses
$\delta\epsilon(L)=\delta\epsilon_\infty+aL^{-q}$.}
\label{fig:energy-fidelity-sm}
\end{figure}

\section{Finite-size Gaussian optimization and numerical critical analysis}
\label{sm:finite-size-numerics}

The thermodynamic Gaussian equations developed in the main Letter provide direct
predictions for the critical line, the magnetic correlation exponent, and
the logarithmic entanglement coefficient. We now test these predictions
independently by optimizing finite-size Gaussian states for the original
long-range transverse-field Ising Hamiltonian.

For each system size $L$, interaction exponent $\alpha$, and transverse
field $h$, we minimize the expectation value of the finite-size Hamiltonian. Using the spin correlators derived in
Sec.~\ref{sm:jw-route}, the variational energy is evaluated directly as
\begin{equation}
E_L[\mathbf R]
=
-\sum_{1\le l<m\le L}
J_{lm}\,
\langle\sigma_l^x\sigma_m^x\rangle_{\mathbf R}
-h\sum_{l=1}^{L}
\langle\sigma_l^z\rangle_{\mathbf R},
\label{eq:finite-variational-energy}
\end{equation}
where the power-law interaction is periodized by summing over all lattice
images,
\begin{equation}
J_{lm}
=
J\sum_{n\in\mathbb Z}
\frac{1}{|l-m+nL|^\alpha}
=
\frac{J}{L^\alpha}
\left[
\zeta\!\left(\alpha,\frac{r_{lm}}{L}\right)
+
\zeta\!\left(\alpha,1-\frac{r_{lm}}{L}\right)
\right],
\label{eq:J-finite}
\end{equation}
with
$r_{lm}=\min(|l-m|,L-|l-m|)$.
At fixed separation, this prescription has the thermodynamic limit
$J_{lm}\to J/r_{lm}^{\alpha}$ as $L\to\infty$.  The optimized state is
\begin{equation}
\mathbf R_\star
=
\underset{\mathbf R}{\operatorname{argmin}}\,
E_L[\mathbf R],
\qquad
|\psi_{\rm var}\rangle=|\mathbf R_\star\rangle .
\label{eq:Rstar-finite}
\end{equation}

The finite-size calculation serves two complementary purposes.  First, for
sizes accessible to exact diagonalization, the optimized Gaussian state can
be compared directly with the exact ground state through its energy,
fidelity, and fermionic antiflatness.  Second, substantially larger systems
can be studied through Gaussian correlation functions, allowing independent
finite-size determinations of the critical field, magnetic exponent, and
logarithmic entanglement coefficient.  We expect these quantities to
approach, with increasing $L$, the critical behavior obtained independently
from the thermodynamic Euler equation in the main Letter.

\subsection{Variational accuracy and non-Gaussianity}
\label{sm:diagnostics}

For system sizes accessible to exact diagonalization, we compare the
optimized Gaussian state with the exact ground state
$|\psi_{\rm ex}\rangle$.  The first diagnostic is the relative energy error,
which we define as
\begin{equation}
\delta\epsilon
=
\frac{|E_{\rm var}-E_{\rm ex}|}{|E_{\rm ex}|},
\qquad
E_{\rm var}=E_L[\mathbf R_\star].
\label{eq:relative-energy-error}
\end{equation}
Because the variational state is obtained by minimizing the energy,
$E_{\rm var}$ is an upper bound to the exact ground-state energy.

A more direct wave-function comparison is provided by the fidelity
\begin{equation}
F
=
\left|
\langle\psi_{\rm ex}|\psi_{\rm var}\rangle
\right|^2 ,
\label{eq:fidelity-sm}
\end{equation}
with $0\le F\le1$.  The fidelity measures how closely the optimized
Gaussian wave function reproduces the exact many-body ground state.

\begin{figure}[t]
\centering
\safeincludegraphics[width=0.48\columnwidth]{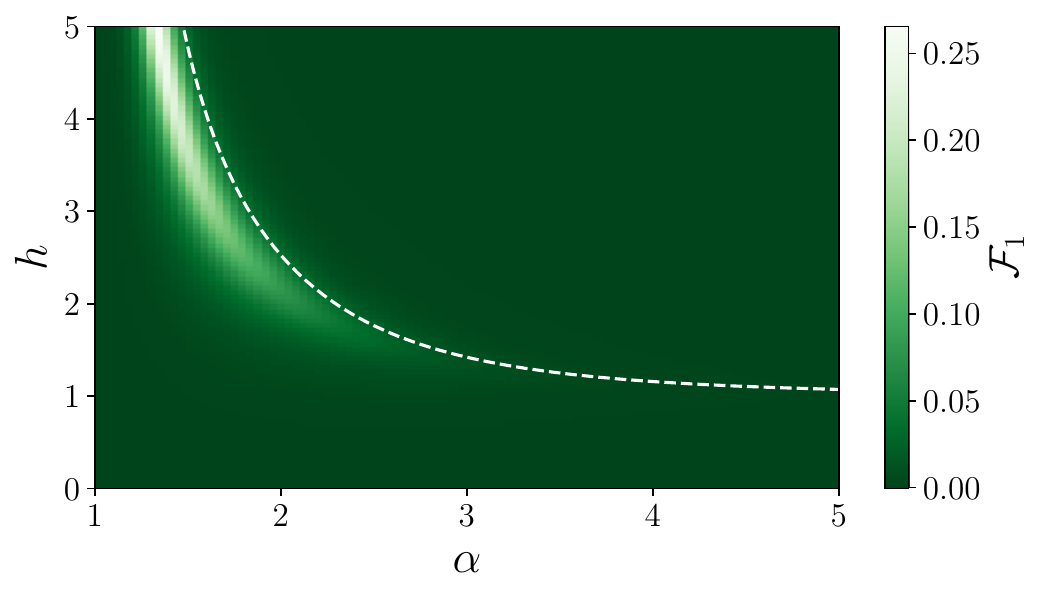}%
\hfill
\safeincludegraphics[width=0.48\columnwidth]{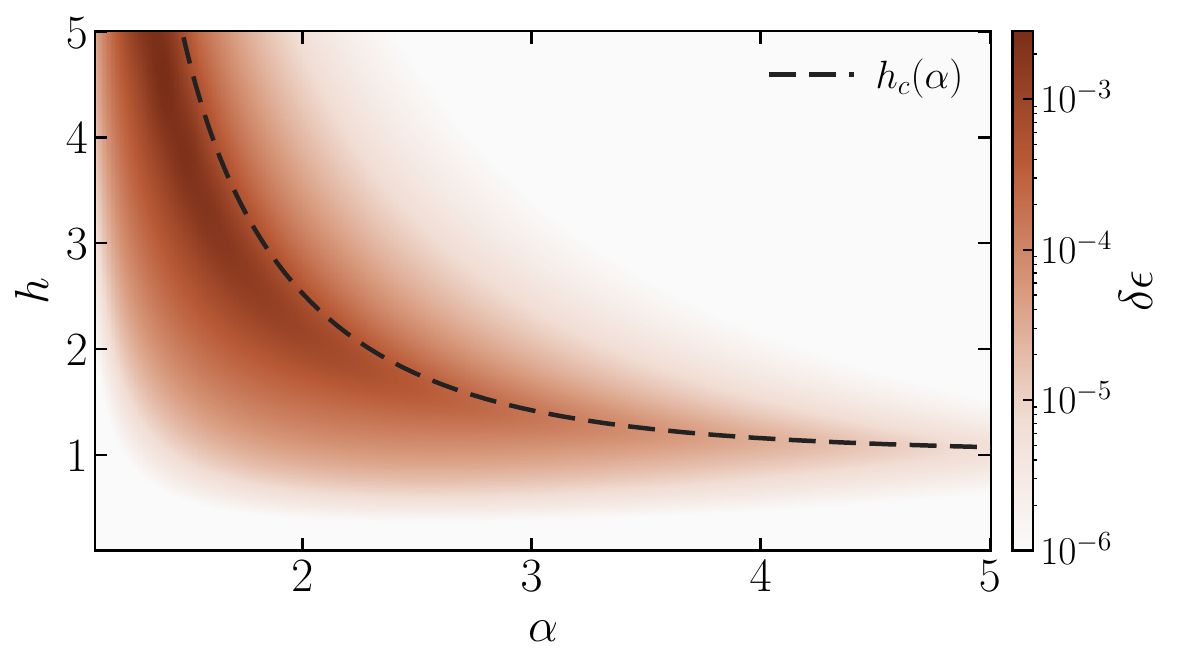}
\caption{\textbf{Non-Gaussianity and Gaussian variational accuracy.}
Left: fermionic antiflatness density of the exact ground state.
Right: relative ground-state energy error $\delta\epsilon$ of the
optimized Gaussian state; the dashed curve marks the thermodynamic
Gaussian critical line.
The strongest departure from Gaussianity and the largest variational
energy errors occur near the transition, while the Gaussian variational
manifold nevertheless yields an accurate critical description.
System sizes and numerical details are specified in the accompanying text.}
\label{fig:anti-flat}
\end{figure}

To quantify the non-Gaussianity of the exact state itself, we use the
fermionic antiflatness introduced in Ref.~\cite{Sierant2026}.  The
Majorana covariance matrix $\boldsymbol\Gamma$ is defined earlier in
Sec.~\ref{sm:full-gaussian}; we therefore do not repeat its construction
here.  The $k$th antiflatness is
\begin{equation}
\mathcal F_k(|\psi\rangle)
=
L-\frac12
\operatorname{Tr}
\left[
(\boldsymbol\Gamma^T\boldsymbol\Gamma)^k
\right].
\label{eq:FAF}
\end{equation}
For a pure Gaussian state,
$\boldsymbol\Gamma^T\boldsymbol\Gamma=\mathbf I_{2L}$, so that
$\mathcal F_k=0$.  A nonzero antiflatness therefore measures the departure
of the target state from the pure Gaussian manifold.

The energy error and fidelity shown in Fig.~\ref{fig:accuracy-main} of the
main Letter indicate that the Gaussian approximation is least accurate in
the vicinity of the transition. Additional finite-size results for the
energy error and fidelity are shown in Fig.~\ref{fig:energy-fidelity-sm},
confirming the same behavior across the sizes considered. To probe the
departure from Gaussianity directly at the level of the exact many-body
state, we additionally examine the fermionic antiflatness in
the left panel of Fig.~\ref{fig:anti-flat}. Its enhancement near the transition
provides finite-size evidence that the exact ground state is most strongly
non-Gaussian in the same region where the Gaussian fidelity is most
strongly reduced. Thus the accuracy of the Gaussian variational predictions does
not rely on the exact critical state becoming Gaussian.
\subsection{Entanglement peak and finite-size critical field}
\label{sm:entropy-peak}

The half-chain von Neumann entropy provides a first independent finite-size
estimate of the critical field. For each optimized Gaussian state, we
restrict the Majorana covariance matrix $\boldsymbol{\Gamma}$ to a block of
$\ell=L/2$ consecutive sites. Denoting the eigenvalues of the resulting
$2\ell\times2\ell$ matrix $\boldsymbol{\Gamma}_\ell$ by
$\{\pm i\nu_j\}_{j=1}^{\ell}$, with $0\leq\nu_j\leq1$, the von Neumann
entropy is
\begin{equation}
S_1(\ell)
=
-\sum_{j=1}^{\ell}
\left[
\frac{1+\nu_j}{2}
\log\frac{1+\nu_j}{2}
+
\frac{1-\nu_j}{2}
\log\frac{1-\nu_j}{2}
\right].
\label{eq:entropy-finite}
\end{equation}
We denote the resulting half-chain entropy by $S_1(L/2;h)$.

For fixed $(L,\alpha)$, the half-chain entropy develops a pronounced
enhancement in the vicinity of the transition.  We define the location of
its maximum by
\begin{equation}
h_{\rm peak}(L,\alpha)
=
\underset{h}{\operatorname{argmax}}\,
S_1(L/2;h).
\label{eq:hpeak-def}
\end{equation}
The finite-size drift of this maximum is then extrapolated to obtain
\begin{equation}
h_c^{(\rm EE)}(\alpha)
=
\lim_{L\rightarrow\infty}
h_{\rm peak}(L,\alpha).
\label{eq:hc-entropy}
\end{equation}
These extrapolated entanglement-peak positions are one of the two
independent finite-size estimates of the critical line used in
Fig.~\ref{fig:hc_exponent_main} of the main Letter.

\begin{figure}[t]
\centering

\safeincludegraphics[width=0.31\linewidth]
{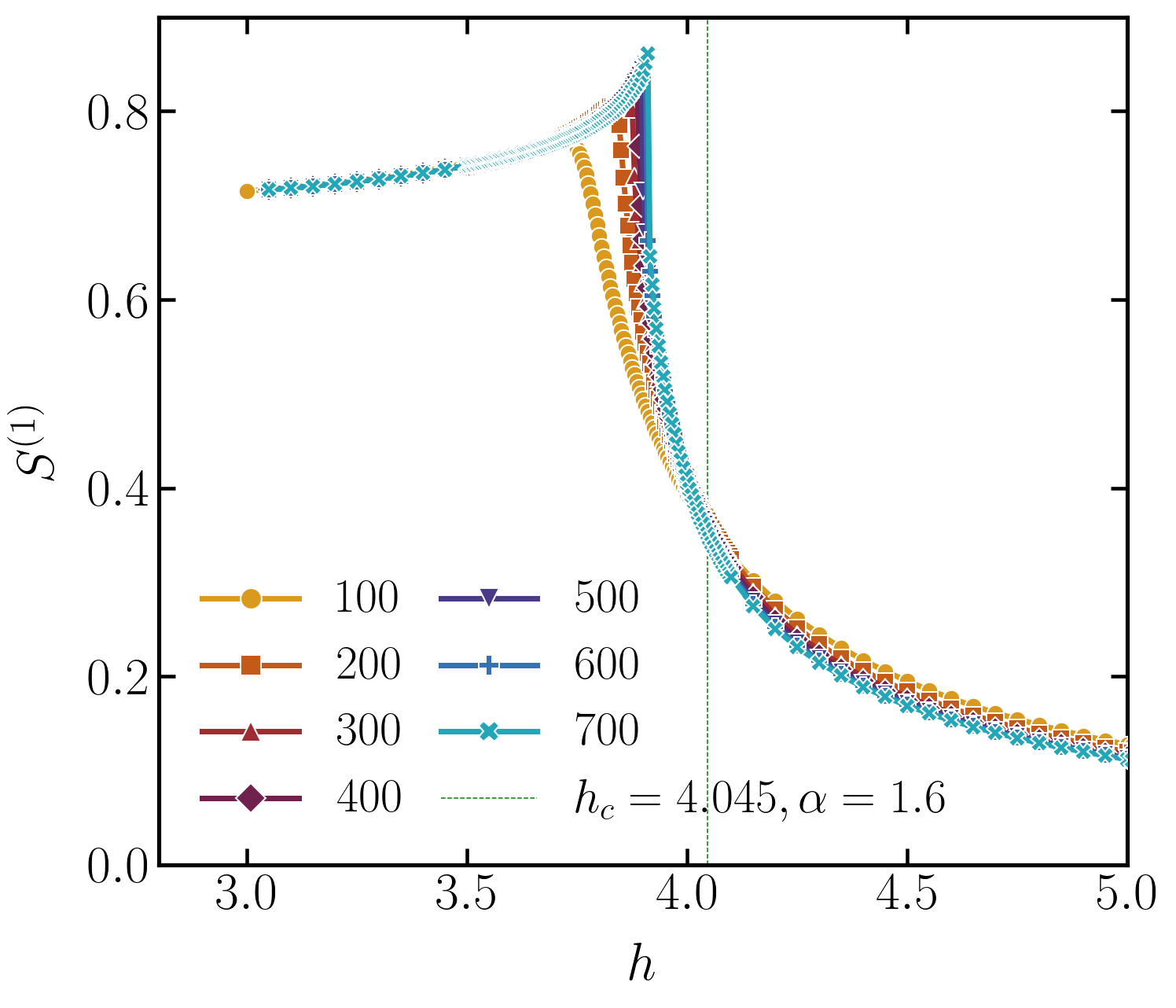}
\safeincludegraphics[width=0.31\linewidth]
{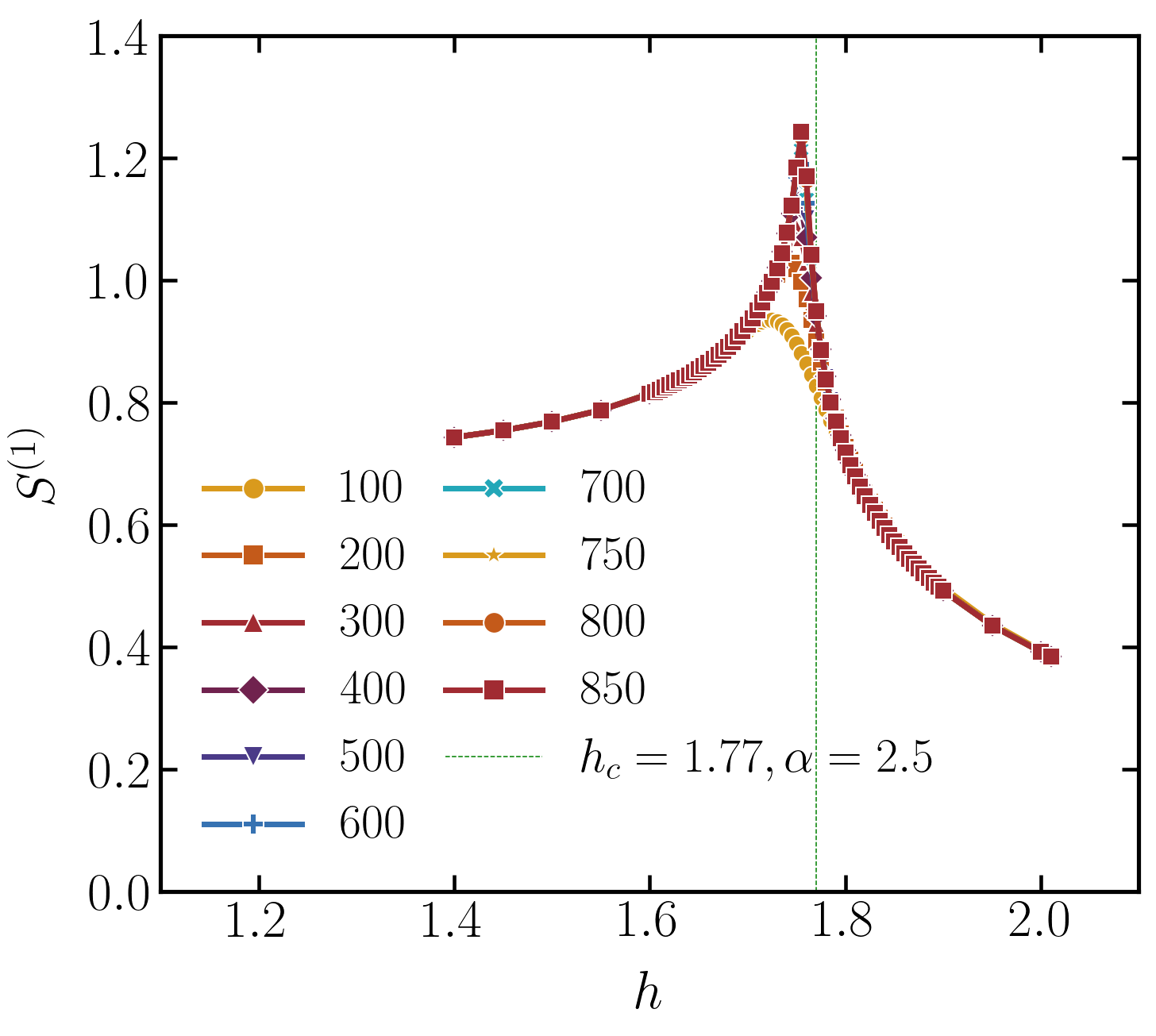}
\safeincludegraphics[width=0.31\linewidth]
{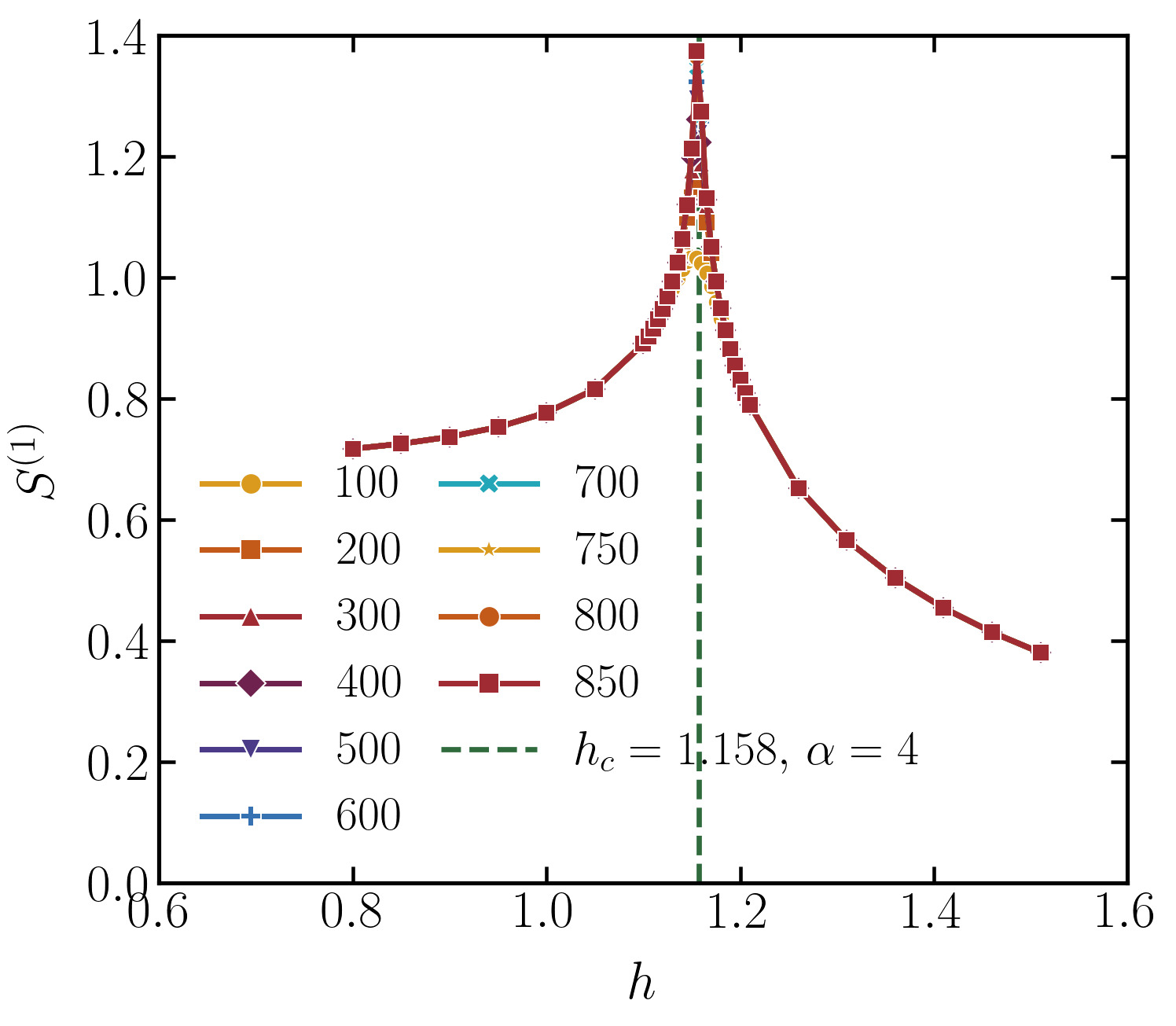}

\caption{\textbf{Half-chain entanglement across the transition.}
Von Neumann entropy $S_1(L/2)$ as a function of the transverse field for
several system sizes at
(a) $\alpha=1.6$, (b) $\alpha=2.5$, and (c) $\alpha=4$.
The finite-size positions of the entropy maxima are extrapolated to obtain
$h_c^{(\rm EE)}(\alpha)$.}
\label{fig:entropy-multi-alpha-sm}
\end{figure}

Representative entropy scans are shown in
Fig.~\ref{fig:entropy-multi-alpha-sm}.  A common feature is the enhancement
of $S_1(L/2)$ close to the critical field, followed by reduced entanglement
as the system moves farther into either phase.  The detailed finite-size
shape depends appreciably on the range of the interaction.

For $\alpha=4$, the entropy develops a relatively sharp cusp-like maximum
close to the critical field, whose height increases with system size.  At
$\alpha=2.5$, the peak remains clearly visible but is broader and more
asymmetric at finite $L$.  For the more strongly long-range case
$\alpha=1.6$, finite-size effects are substantially stronger: the entropy
maximum is broad, its position drifts visibly with $L$, and the descent on
the paramagnetic side sharpens only gradually with increasing size.
Consequently, extrapolation of $h_{\rm peak}(L,\alpha)$ is particularly
important in the long-range regime.

These entropy scans therefore play two distinct roles in the analysis:
their peak positions provide the finite-size critical-field estimates used
in the main critical-line figure, while their values at the extrapolated
critical point are used below to determine the logarithmic entanglement
coefficient.

\subsection{Binder-cumulant estimate of the critical field}
\label{sm:binder-crossings}

A second and independent finite-size determination of the transition is
obtained from the Binder cumulant.  We define the extensive and intensive
magnetizations along the ordering direction by
\begin{equation}
M_x=\sum_{j=1}^{L}\sigma_j^x,
\qquad
m_x=\frac{M_x}{L},
\label{eq:mx-def}
\end{equation}
and the Binder cumulant by
\begin{equation}
U_L
=
1-
\frac{
\langle m_x^4\rangle
}{
3\langle m_x^2\rangle^2
}.
\label{eq:Binder_def}
\end{equation}
The second and fourth moments entering Eq.~\eqref{eq:Binder_def} are
computed using the two- and four-spin correlation functions derived in
Sec.~\ref{sm:jw-route}.

\begin{figure}[t]
\centering

\safeincludegraphics[width=0.35\linewidth]
{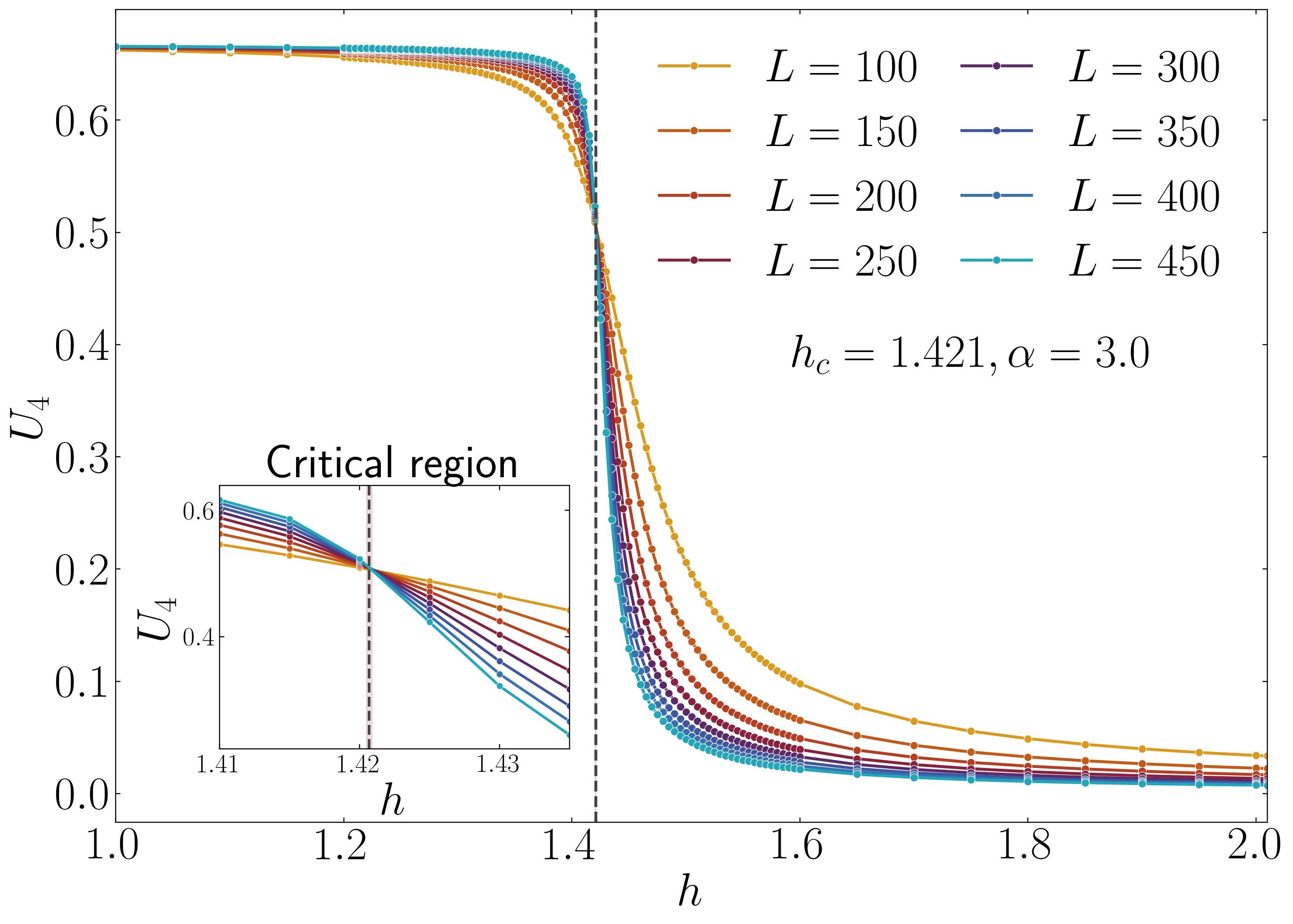}
\safeincludegraphics[width=0.35\linewidth]
{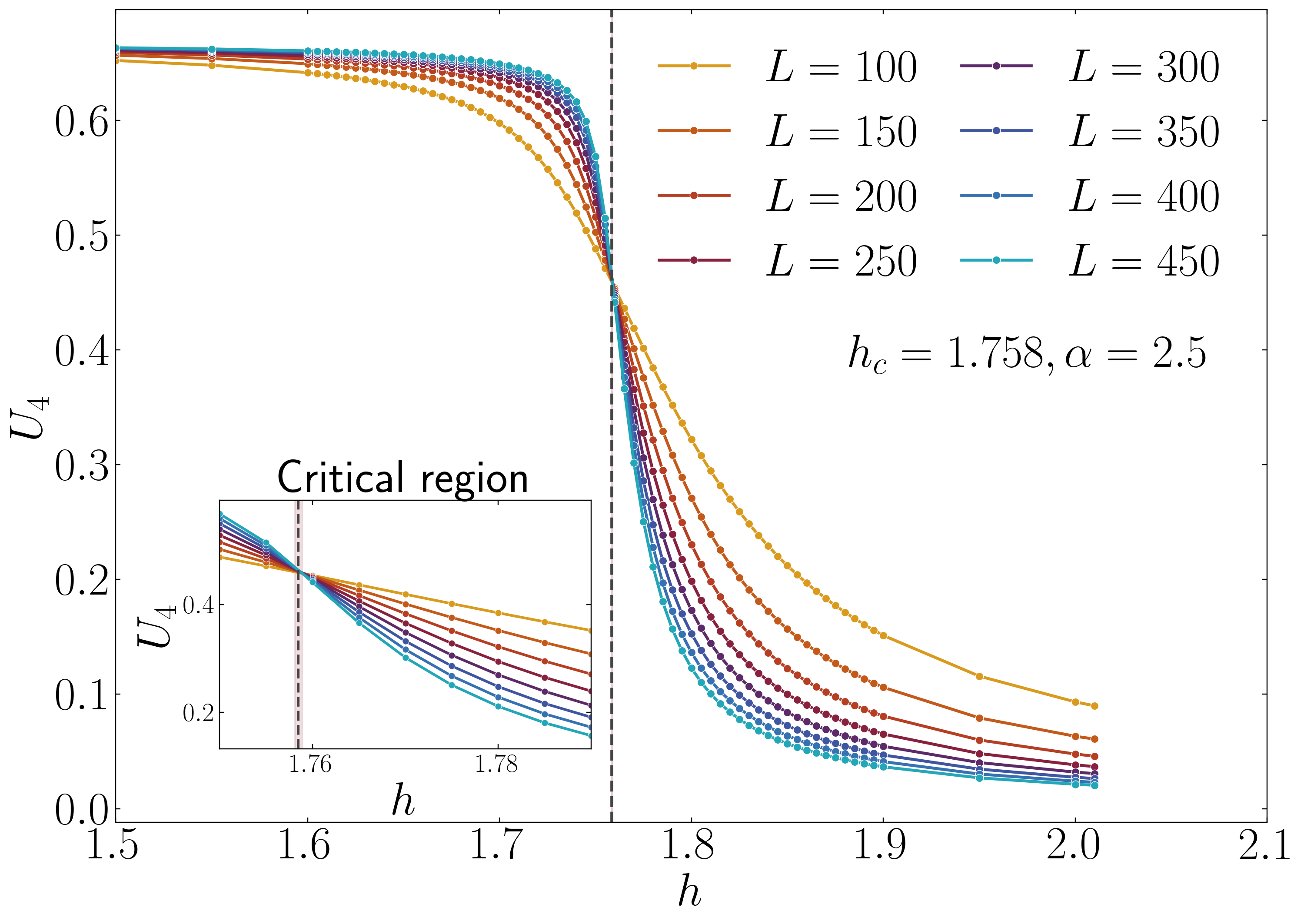}
\safeincludegraphics[width=0.35\linewidth]
{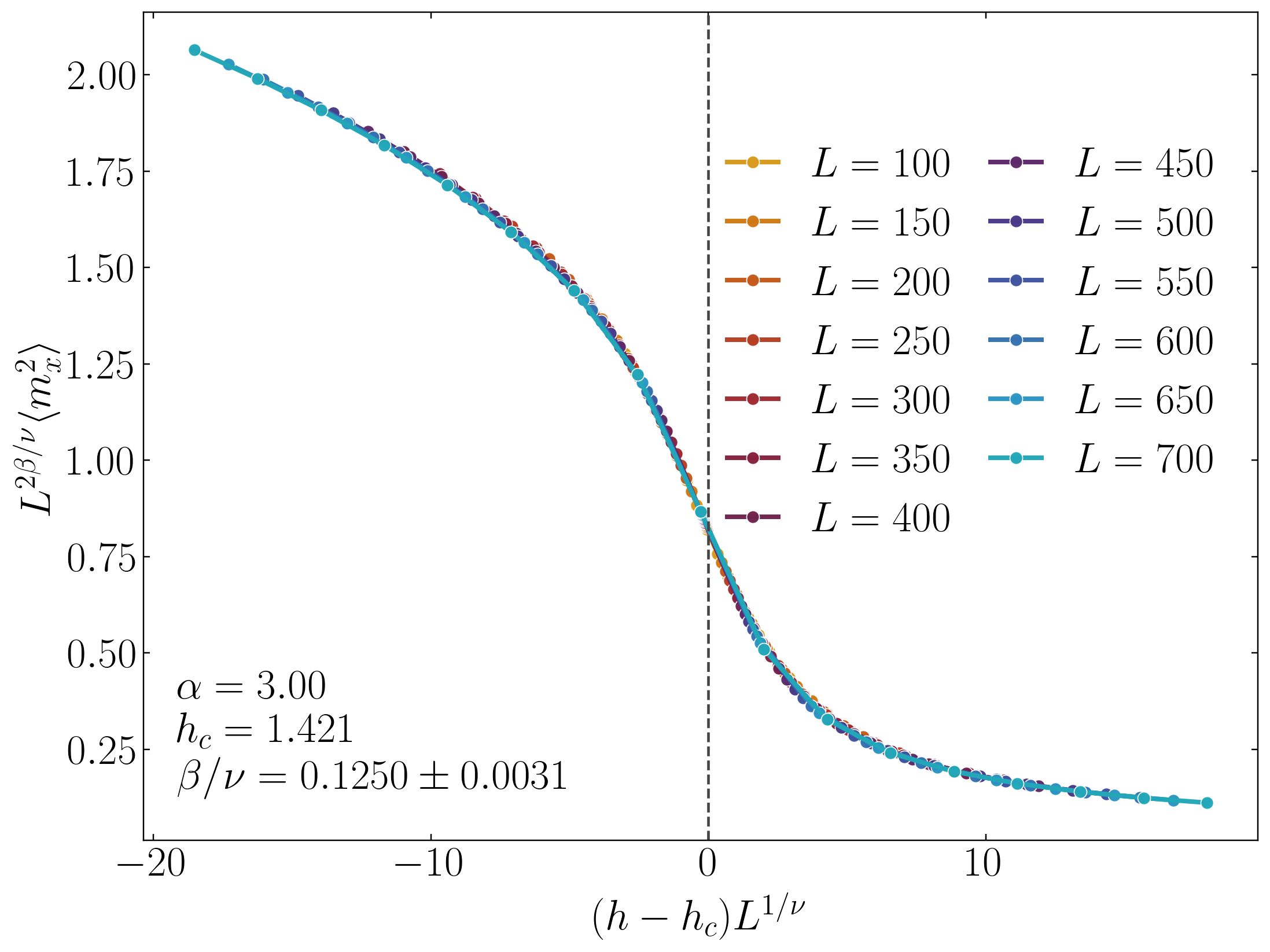}
\safeincludegraphics[width=0.35\linewidth]
{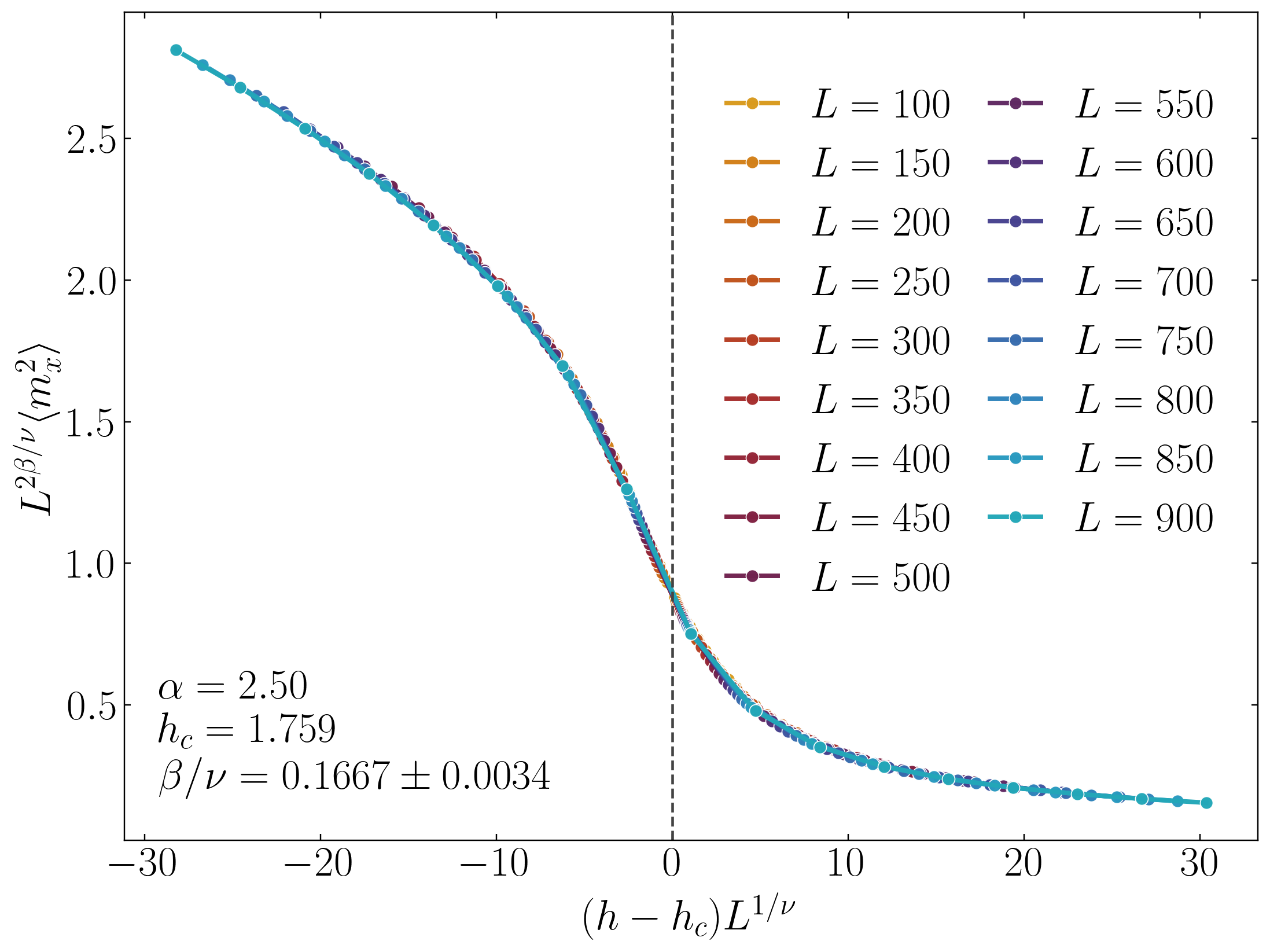}

\caption{\textbf{Binder-cumulant crossings and magnetic finite-size scaling.}
\textbf{Upper panels:} Representative Binder cumulants for several system
sizes at $\alpha=3$ and $\alpha=2.5$. The size-dependent crossings determine
$h_{\rm cross}(L,\alpha)$, whose finite-size drift is extrapolated to obtain
$h_c^{(\rm B)}(\alpha)$.
\textbf{Lower panels:} Representative finite-size collapses of
$\langle m_x^2\rangle$ for $\alpha=3$ and $\alpha=2.5$, using the scaling
form in Eq.~\eqref{eq:mx2_scaling}.}
\label{fig:binder-collapses-examples}
\end{figure}

Because $U_L$ is dimensionless, its critical behavior is less sensitive to
nonuniversal amplitudes than that of the order parameter itself.  Near the
transition, curves for different system sizes become weakly size dependent
and exhibit crossings in the vicinity of $h_c$.  From the crossing of
successive sizes we obtain a size-dependent estimate
$h_{\rm cross}(L,\alpha)$ and extrapolate its drift according to
\begin{equation}
h_c^{(\rm B)}(\alpha)
=
\lim_{L\rightarrow\infty}
h_{\rm cross}(L,\alpha).
\label{eq:hc-binder}
\end{equation}
The values $h_c^{(\rm B)}(\alpha)$ constitute the second finite-size data
set compared with the direct thermodynamic Gaussian critical line in
Fig.~\ref{fig:hc_exponent_main}. Representative Binder-cumulant curves are shown in
Fig.~\ref{fig:binder-collapses-examples}.

\subsection{Magnetic finite-size scaling and exponent extraction}
\label{sm:fss}
\label{subsec:critical}

The magnetic critical exponent is obtained independently from the
finite-size behavior of the squared order parameter
\begin{equation}
\langle m_x^2\rangle
=
\frac{1}{L^2}
\sum_{i,j=1}^{L}
\langle\sigma_i^x\sigma_j^x\rangle .
\label{eq:mx2_def}
\end{equation}
For the long-range model it is useful to write the finite-size scaling form
in terms of the generalized scaling dimensions
\begin{equation}
x_m
\equiv
\frac{\beta}{\nu},
\qquad
y_h
\equiv
\frac{1}{\nu},
\label{eq:generalized-exponents}
\end{equation}
The scaling forms are
\begin{align}
\langle m_x^2\rangle(h)
&=
L^{-2x_m}
\mathcal M
\left[
(h-h_c)L^{y_h}
\right],
\label{eq:mx2_scaling}
\\
U_L(h)
&=
\mathcal U
\left[
(h-h_c)L^{y_h}
\right].
\label{eq:Binder_scaling}
\end{align}
When ordinary finite-size scaling applies, $\theta=1$, so that
$x_m=\beta/\nu$ and $y_h=1/\nu$.  Operationally, the Binder crossings provide an initial estimate of $h_c$.
The exponent $y_h$ is constrained by the critical scaling region, while
$x_m$ is determined from the collapse and from the size dependence of
$\langle m_x^2\rangle$ at criticality.

For a trial set $(h_c,y_h,x_m)$, the rescaled variables are
\begin{equation}
x=(h-h_c)L^{y_h},
\qquad
y=L^{2x_m}\langle m_x^2\rangle.
\label{eq:fss-rescaled}
\end{equation}
Representative collapses are displayed in
Fig.~\ref{fig:binder-collapses-examples}.  For the values shown there,
$\alpha=2.5,3,$ and $4$, ordinary finite-size scaling applies and hence
$x_m=\beta/\nu$.  At $\alpha=4$ we obtain
$\beta/\nu\simeq0.125$ and $h_c\simeq1.156$.
For $\alpha=3$ we similarly find
$\beta/\nu\simeq0.125$ and $h_c\simeq1.420$.
At $\alpha=2.5$, the critical field shifts to
$h_c\simeq1.759$ and the magnetic exponent to approximately
$\beta/\nu\simeq0.1667$, displaying the continuously varying
long-range behavior.

The finite-size magnetic exponents obtained in this way provide the
numerical Gaussian points compared with the analytic prediction in
Fig.~\ref{fig:hc_exponent_main}(a) of the main Letter.

\begin{figure}[t]
\centering

\safeincludegraphics[width=0.4\linewidth]
{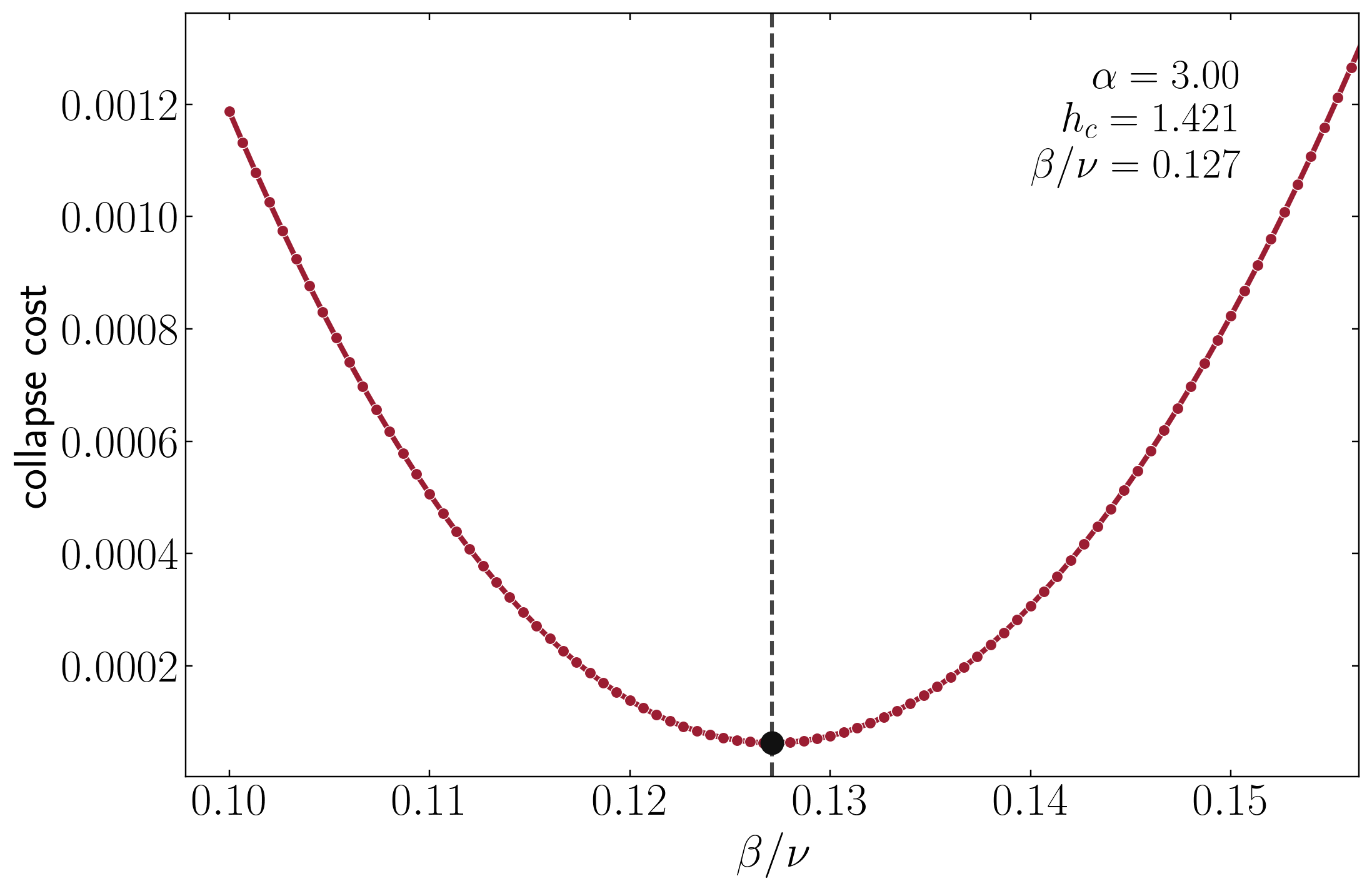}
\safeincludegraphics[width=0.4\linewidth]
{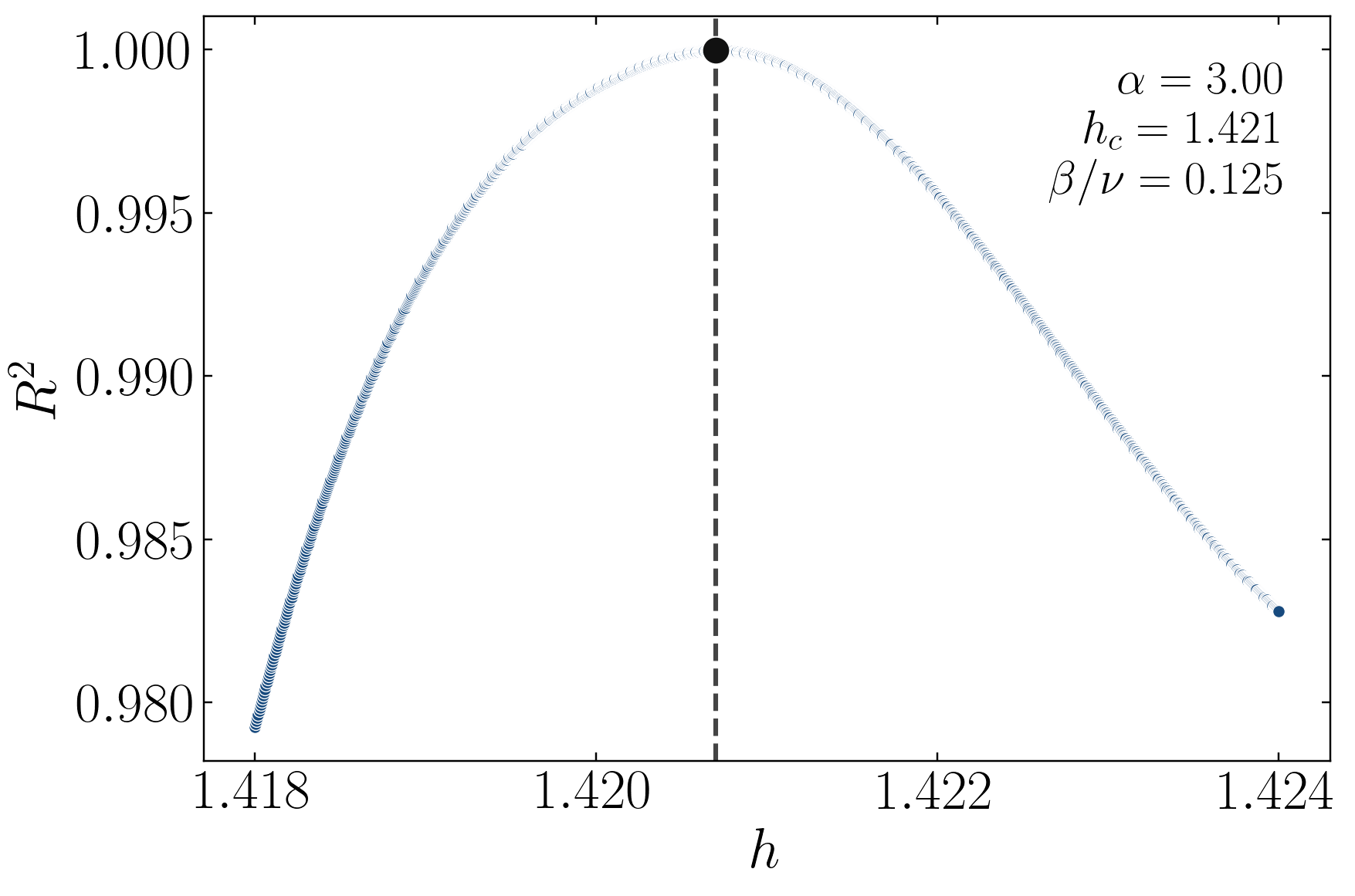}

\caption{\textbf{Stability of the magnetic scaling analysis.}
Representative collapse-cost and $R^2$ diagnostics used to determine the
critical parameters and select the fitting window.}
\label{fig:Cost_R2}
\end{figure}

\paragraph{Data-collapse estimator.}
A practical issue in the scaling analysis is the choice of the field
interval $[h_{\min},h_{\max}]$.  If the interval is too narrow, the fit
becomes noisy and unstable, while an excessively wide interval introduces
noncritical corrections.  We therefore repeat the analysis for several
candidate windows.

For each trial window and each set $(h_c,y_h,x_m)$, the rescaled curves
$y(x)$ are interpolated onto a common grid of $N_x$ points in their overlap
region.  Denoting by $y_L(x_k)$ the interpolated curve for size $L$, we
define
\begin{equation}
\overline y(x_k)
=
\frac{1}{N_L}
\sum_L y_L(x_k)
\end{equation}
and use the normalized collapse cost
\begin{equation}
C(h_c,y_h,x_m)
=
\frac{1}{N_x}
\sum_{k=1}^{N_x}
\frac{
\displaystyle
\frac{1}{N_L}
\sum_L
\left[
y_L(x_k)-\overline y(x_k)
\right]^2
}{
\overline y(x_k)^2
}.
\label{eq:collapse-cost}
\end{equation}
The optimal parameters minimize $C$, and the preferred field window is the
one producing the smallest stable collapse cost.

\paragraph{$R^2$ estimator.}
As an independent estimate, we analyze the size dependence of
$\langle m_x^2\rangle$ at fixed $h$.  For every field value we fit
\begin{equation}
\langle m_x^2\rangle
\sim
L^{s(h)},
\end{equation}
or equivalently
\begin{equation}
\log\langle m_x^2\rangle
=
a(h)+s(h)\log L.
\label{eq:log-mx2-fit}
\end{equation}
Writing
$y_i=\log\langle m_x^2\rangle_i$ and
$\hat y_i=a(h)+s(h)\log L_i$, the quality of the fit is measured by
\begin{equation}
R^2(h)
=
1-
\frac{
\displaystyle
\sum_{i=1}^{N_L}(y_i-\hat y_i)^2
}{
\displaystyle
\sum_{i=1}^{N_L}(y_i-\bar y)^2
},
\qquad
\bar y
=
\frac{1}{N_L}
\sum_{i=1}^{N_L}y_i .
\label{eq:R2-def}
\end{equation}
The critical field is identified close to the maximum of $R^2(h)$, and
the corresponding magnetic scaling dimension follows from
\begin{equation}
x_m=-\frac{s(h_c)}{2}.
\label{eq:xm-R2}
\end{equation}
Thus the preferred fitting interval for this method is the one producing
the largest stable value of $R^2$. A representative example of the collapse-cost and $R^2$ diagnostics is
shown in Fig.~\ref{fig:Cost_R2}.

\begin{figure}[t]
\centering
\safeincludegraphics[width=0.4\linewidth]
{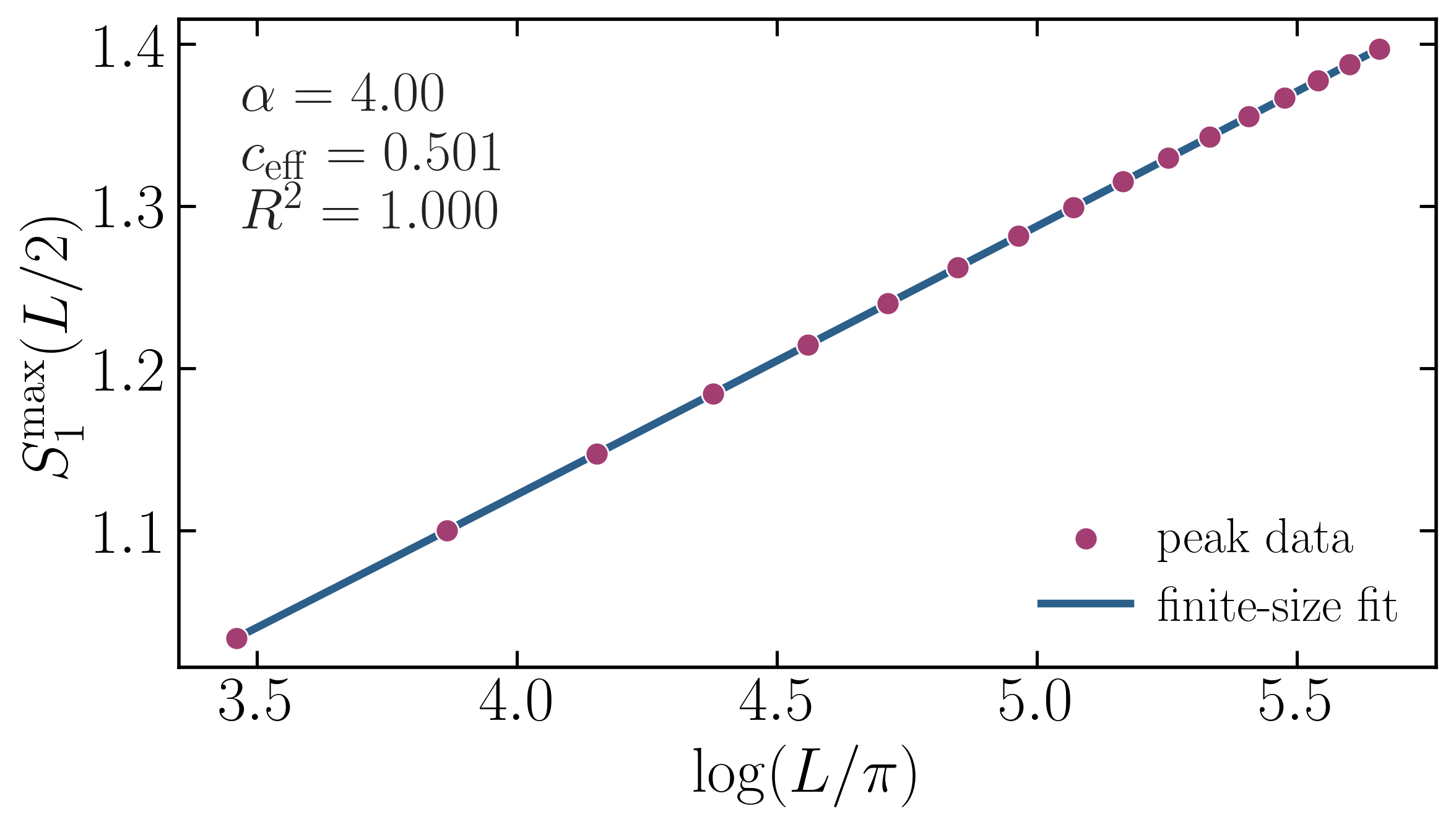}
\safeincludegraphics[width=0.4\linewidth]
{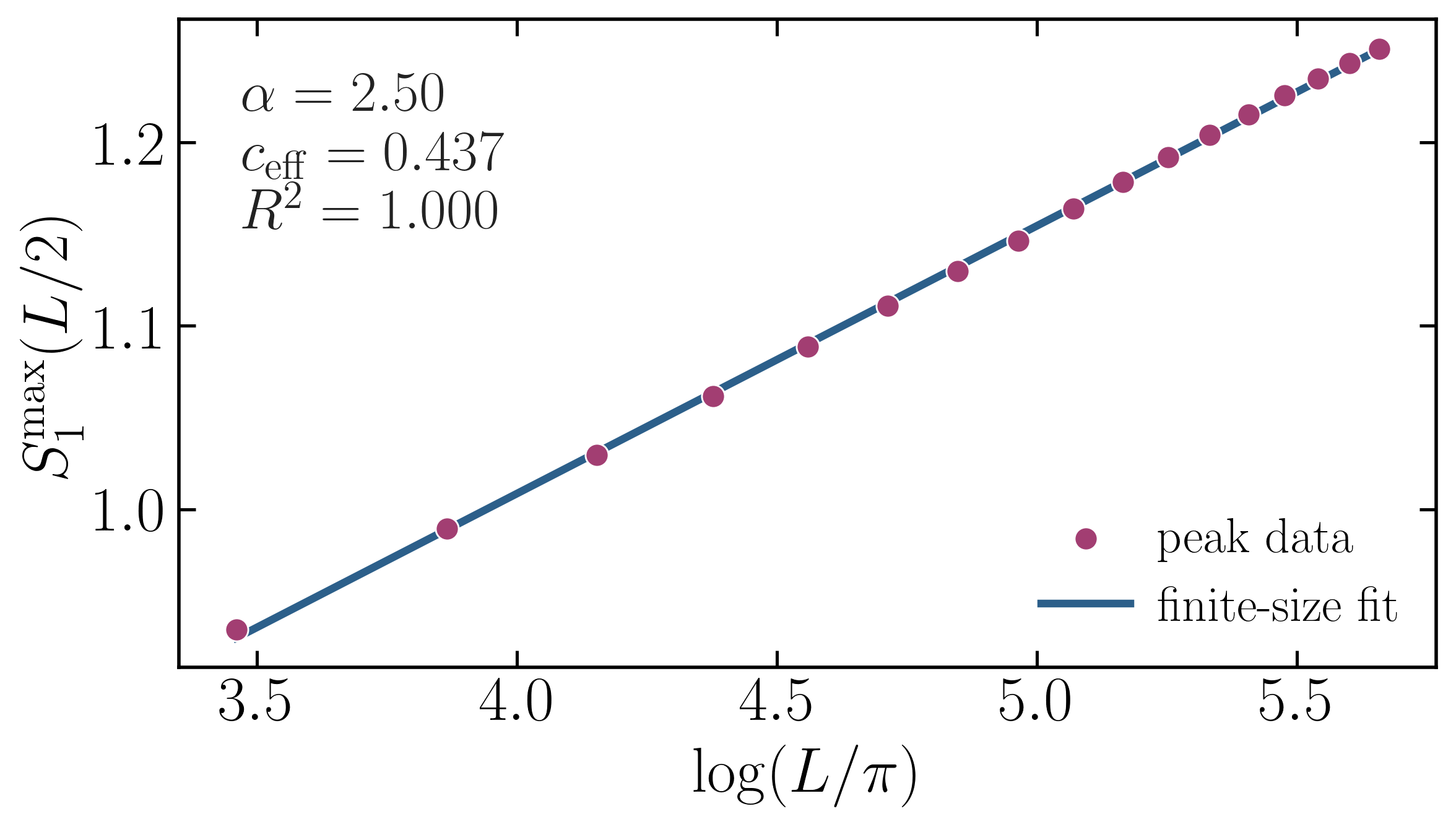}
\caption{\textbf{Extraction of the logarithmic entanglement coefficient.}
Representative fits of the critical half-chain entropy
$S_1(L/2)$ versus $\log L$.  According to
Eq.~\eqref{eq:ceff-halfchain-fit}, the fitted slope
$A(\alpha)$ determines $c_{\rm eff}(\alpha)=3A(\alpha)$.}
\label{fig:ceff-fits-sm}
\end{figure}

\subsection{Entanglement scaling and effective central charge}
\label{sm:ceff-numerics}
\label{subsec:entropy}

We finally determine numerically the logarithmic entanglement coefficient
$c_{\rm eff}(\alpha)$ compared with the analytic Gaussian prediction in
Fig.~\ref{fig:ceff-main}.  This numerical extraction uses only the
finite-size optimized Gaussian states and is independent of the
dilogarithm expression derived in the End Matter.

For a subsystem $A$, the covariance matrix is restricted to the
$2|A|$ Majorana modes contained in $A$.  If its eigenvalues are
$\pm\ii\nu_r$, with $0\le\nu_r\le1$, we define
\begin{equation}
p_r^\pm=\frac{1\pm\nu_r}{2}.
\end{equation}
The Gaussian Rényi entropy is
\begin{equation}
S_q(A)
=
\frac{1}{1-q}
\sum_r
\log
\left[
(p_r^+)^q+(p_r^-)^q
\right],
\qquad q\neq1,
\label{eq:Renyi-Gaussian-sm}
\end{equation}
while its $q\to1$ limit gives the von Neumann entropy used throughout the
main Letter [Eq.~\eqref{eq:entropy-finite}].

For a periodic critical chain described by a local conformal field theory,
the block entropy obeys
\begin{equation}
S_1(\ell)
=
\frac{c}{3}
\log d(\ell)+B,
\qquad
d(\ell)
=
\frac{L}{\pi}
\sin\left(\frac{\pi\ell}{L}\right),
\label{eq:EE_chord}
\end{equation}
where $d(\ell)$ is the chord distance and $B$ is nonuniversal.  We use this
form as a numerical characterization of the logarithmic entropy growth.

There are two complementary finite-size extractions.  First, for fixed
$L$ we fit the block-size dependence to
\begin{equation}
S_1(\ell)
=
A\log d(\ell)+B,
\label{eq:block-entropy-fit}
\end{equation}
which gives
\begin{equation}
c_{\rm eff}=3A.
\label{eq:ceff-block}
\end{equation}
This provides a useful check of the logarithmic form and of finite-size
effects.

The main numerical values used in Fig.~\ref{fig:ceff-main}, however, are
obtained from the half-chain entropy.  Setting $\ell=L/2$ in
Eq.~\eqref{eq:EE_chord} gives
\begin{equation}
S_1(L/2)
=
\frac{c}{3}\log L+B',
\label{eq:halfchain-CFT}
\end{equation}
where the additive term $-(c/3)\log\pi$ has been absorbed into $B'$.
We therefore fit the optimized critical Gaussian data to
\begin{equation}
S_1(L/2)
=
A(\alpha)\log L+B(\alpha),
\qquad
c_{\rm eff}(\alpha)=3A(\alpha).
\label{eq:ceff-halfchain-fit}
\end{equation}
The entropy entering this fit is evaluated at the corresponding critical
field $h_c(\alpha)$.  Representative half-chain fits will be shown in
Fig.~\ref{fig:ceff-fits-sm}.

As an additional consistency check, the Rényi entropies may be analyzed
using the local-CFT form
\begin{equation}
S_q(\ell)
=
\frac{c}{6}
\left(1+\frac{1}{q}\right)
\log d(\ell)+B_q.
\label{eq:Renyi-CFT}
\end{equation}
If
$S_q(\ell)=A_q\log d(\ell)+B_q$, this gives
\begin{equation}
c_{\rm eff}^{(q)}
=
\frac{6q}{q+1}A_q .
\label{eq:ceff-Renyi}
\end{equation}
In a conventional conformal regime these estimates should be independent
of $q$ and agree with the von Neumann result.

For sufficiently short-range interactions the extracted coefficient
approaches the Ising value $c=1/2$.  For smaller $\alpha$ it varies with
the interaction range.  In this genuinely long-range regime we do not
assume that the coefficient is a Virasoro central charge; we therefore use
$c_{\rm eff}(\alpha)$ as the effective coefficient of logarithmic
entanglement growth.  The half-chain values obtained from
Eq.~\eqref{eq:ceff-halfchain-fit} are the numerical points compared with
the analytic Gaussian prediction in Fig.~\ref{fig:ceff-main}.

\end{document}